\documentclass[prd,floats,superscriptaddress,showpacs,nofootinbib,preprintnumbers]{revtex4}

\usepackage{amssymb}
\usepackage{amsmath}

\usepackage{sans}

\usepackage{graphicx}
\usepackage{graphics}
\usepackage{dcolumn}
\usepackage{color}
\usepackage{rotate}
\usepackage{fancyhdr}
\usepackage{hyperref}
\usepackage{indentfirst}

\usepackage{umoline}
\usepackage{ulem}

\def\beq{\begin{equation}}
\def\eeq{\end{equation}}
\def\be{\begin{eqnarray}}
\def\ee{\end{eqnarray}}
\def\ba{\begin{eqnarray}}
\def\ea{\end{eqnarray}}
\def\no{\nonumber}

\def\C{\mathcal{C}}

\def\F{\mathcal{F}}
\def\D{\mathcal{D}}

\def\B{\mathcal{B}}

\definecolor{darkred}{rgb}{.743,0,0}

\def\n02b{$0\nu\beta\beta$}
\def\n02bphi{$0\nu\beta\neta\phi$}
\def\lsim{\mathrel{\rlap{\lower4pt\hbox{\hskip1pt$\sim$}}
    \raise1pt\hbox{$<$}}}         
\def\gsim{\mathrel{\rlap{\lower4pt\hbox{\hskip1pt$\sim$}}
    \raise1pt\hbox{$>$}}}         

\begin{document}
\preprint{CERN-TH-2020-110}

\title{On coalescence as the origin of nuclei in hadronic collisions}

\author{Francesca Bellini}
\email{Francesca.Bellini@cern.ch}
\affiliation{Experimental Physics Department, CERN, CH-1211 Geneve 23, Switzerland}
\author{Kfir Blum}
\email{kfir.blum@weizmann.ac.il}
\affiliation{Weizmann Institute, Department of Particle Physics and Astrophysics, Rehovot 7610001, Israel}
\affiliation{Theoretical Physics Department, CERN, CH-1211 Geneve 23, Switzerland}
\author{Alexander Phillip Kalweit}
\email{alexander.philipp.kalweit@cern.ch}
\affiliation{Experimental Physics Department, CERN, CH-1211 Geneve 23, Switzerland}
\author{Maximiliano Puccio}
\email{maximiliano.puccio@cern.ch}
\affiliation{Experimental Physics Department, CERN, CH-1211 Geneve 23, Switzerland}

\begin{abstract}
The origin of weakly-bound nuclear clusters in hadronic collisions is a key question to be addressed by heavy-ion collision (HIC) experiments. The measured yields of clusters are approximately consistent with expectations from   phenomenological statistical hadronisation models (SHMs), but a theoretical understanding of the dynamics of cluster formation prior to kinetic freeze out is lacking. The competing model is nuclear coalescence, which attributes cluster formation to the effect of final state interactions (FSI) during the propagation of the nuclei from kinetic freeze out to the observer. This phenomenon is closely related to the effect of FSI in imprinting femtoscopic correlations between continuum pairs of particles at small relative momentum difference. 
We give a concise theoretical derivation of the coalescence--correlation relation, predicting nuclear cluster spectra from femtoscopic measurements. We review the fact that coalescence derives from a relativistic Bethe-Salpeter equation, and recall how effective quantum mechanics controls the dynamics of cluster particles that are nonrelativistic in the cluster centre of mass frame. We demonstrate that the coalescence--correlation relation is roughly consistent with the observed cluster spectra in systems ranging from PbPb to pPb and pp collisions. Paying special attention to nuclear wave functions, we derive the coalescence prediction for hypertriton and show that it, too, is roughly consistent with the data. Our work motivates a combined experimental programme addressing  femtoscopy and cluster production under a unified framework. Upcoming pp, pPb and peripheral PbPb data analysed within such a programme could stringently test coalescence as the origin of clusters.
\end{abstract}

\maketitle
\tableofcontents

\section{Introduction}
Loosely-bound nuclei like the D, $^3$He, $^3$H, $^3_\Lambda$H and their anti-particles are detected among the products of high-energy hadronic collisions at the large hadron collider (LHC) and other experiments, and their study is a central objective in heavy-ion collision (HIC) experiments~\cite{Csernai:1986qf,Andronic:2010qu,Cleymans:2011pe}. Interestingly, the momentum-integrated yields of these nuclei are roughly consistent with being drawn from a thermal distribution with the same temperature parameter $T_{ch}$ that fits the yields of mesons and nucleons~\cite{Andronic:2017pug,Vovchenko:2018fiy,Vovchenko:2019kes,Acharya:2017bso}. Taken together from $\pi^\pm$ to $^4$He, the hadron yields span 
 some $\sim9$ orders of magnitude with only $\mathcal{O}(1)$ discrepancies\footnote{With the discrepancies affecting nuclei~\cite{Vovchenko:2018fiy} at a comparable level to mesons and nucleons~\cite{Vovchenko:2019kes}.}. This has led some authors to speculate that nuclei take part, on equal footing with the ``more fundamental" mesons and nucleons, in an equilibrium partition function characterising the high excitation state (HXS) produced in HICs. A recent account of this statistical hadronisation model (SHM) is given in~\cite{Andronic:2017pug}. 
 
While the SHM is approximately consistent with cluster yields, no first-principle theoretical framework as of yet explains the dynamics of cluster formation in the HXS before kinetic freeze out\footnote{A recent speculation is the formation of compact ``preclusters"~\cite{Shuryak:2018lgd,Shuryak:2019ikv} due to an in-medium modification of the nuclear potential.}. Clusters are big (several fm) fragile (binding energies $E_b\lesssim10$~MeV, as low as $E_b\sim0.13$~MeV for $^3_\Lambda$H \cite{Juric:1973zq}) objects, while even at kinetic freeze out the HXS does not exceed a few\footnote{By this we mainly have in mind the HXS homogeneity radius as revealed by femtoscopy~\cite{Lisa:2005dd,Heinz:1999rw}, to be discussed later on. But even the total HXS volume as fitted in the SHM in high multiplicity PbPb collisions is only of diameter $\sim20$~fm, shrinking to $\sim2$~fm in low multiplicity pp and pPb collisions~\cite{Vovchenko:2019kes}.} fm and it is a hot state with characteristic particle excitation energies of $\sim100$~MeV. What does it mean for a D with diameter $\sim4$~fm to exist in an equilibrium distribution in an HXS of diameter $\sim2$~fm, produced in pp collisions? What does it mean for $^3_\Lambda$H, with effective diameter $\sim14$~fm? This puzzle makes the origin of nuclei uniquely interesting. 

The kinetic theory analysis of~\cite{Oliinychenko:2018ugs}, focusing on D formation in high-multiplicity PbPb collisions, may shed some light on the problem. This analysis demonstrated that, indeed, D observed at the detector must emerge from the kinetic freeze out region and not from the chemical freeze out region of the HXS, to which the SHM parameter $T_{ch}$ corresponds. However, Ref.~\cite{Oliinychenko:2018ugs} treated the D as a point-particle and it is far from clear if and how their analysis could be adapted to smaller systems like pp, pPb or peripheral PbPb collisions\footnote{We thank Urs Wiedemann for pointing out this issue during a workshop at CERN.}.

An alternative explanation for the origin of nuclei, bypassing the limitations of kinetic theory, is proposed by coalescence. The basic assumption of the coalescence model is that the expansion of the HXS leads to kinetic freeze out with nucleons but -- due to their fragility and size -- essentially no nuclei. The HXS at kinetic freeze out can be described by a quantum mechanical (QM) density matrix. Projecting the density matrix onto particle states at the detector gives the observed particle spectra. Final state interactions (FSI) mediated by nuclear scattering and Coulomb photon exchange enter this projection as they affect the propagation of the particles from the HXS to the detector. 
FSI manifest themselves in two ways:
\begin{enumerate}
\item FSI imprint momentum correlations among pairs of continuum particles. The analysis of this phenomenon is known as femtoscopy or Hanbury Brown-Twiss (HBT) analysis\footnote{We will use the terms femtoscopy and HBT analysis interchangeably in this paper, in line with most of the literature. For a comparative discussion and historical notes, see~\cite{Lednicky:2007ax}.}.
\item FSI also admit discrete bound-state multi-nucleon solutions, namely nuclei. This is nuclear coalescence.
\end{enumerate}

It is important to note that the coalescence model predicts that the yields of nuclei approximately inherit the thermal spectra of their nucleon constituents, up to a dimensionless QM correction factor. In many cases (e.g. low-$p_t$ D and $^3$He formation in high multiplicity PbPb collisions) the QM correction factor is close to unity~\cite{Csernai:1986qf,Scheibl:1998tk}. Thus, the approximately thermal yield of nuclei need not point to the nuclei taking part in an equilibrium partition function before kinetic freeze out. For some nuclei and systems, however, the QM factor is predicted to be much smaller than unity. An example is the production of $^3_\Lambda$H in pp collisions, or (especially at high-$p_t$) in low multiplicity PbPb collisions. These systems offer a key discriminator between the coalescence model and the SHM. Our goal in this paper is to study the theoretical prediction of coalescence, compare to available experimental data and highlight the path to making this test conclusive.

A central feature in this paper is the relation between coalescence and HBT correlations among continuum nucleons. 
From the perspective of the coalescence model, HBT correlations and nuclei production are closely related. In a clear sense, the successful reconstruction of the imprint of FSI on pair correlations lends credence to the basic framework of coalescence, which deals with the bound state solutions of essentially the same FSI (in different isospin channels). Moreover, once HBT calibrates the HXS source characteristics, nuclei yields are predicted without free parameters. 
Over the years the community (experimental and theoretical) developed a habit of considering HBT and nuclei analyses separately, making it cumbersome to combine the information content of the measurements. One of our goals here is to motivate joint experimental analyses of HBT and cluster yields.

The plan of the paper, along with a brief summary, are as follows.

In Sec.~\ref{s:basics} we briefly review the underlying relation between femtoscopy and cluster formation, defining the coalescence/femtoscopy framework. The basic formalism was laid out by Lednicky et al~\cite{Lednicky:1981su,Akkelin:2001nd,Lednicky:2005tb}. We provide a quick reduction of this formalism to observationally accessible objects. This is a good starting point for the discussion, because it demonstrates that coalescence arises in a relativistic quantum field theoretic (QFT) calculation. 
In Sec.~\ref{ss:2ptD} we show how, subject to two key approximations (the smoothness approximation and the equal-time approximation), the model-independent coalescence--correlation relation between deuteron production and two-proton HBT comes about. 
The main result here is the manifest relation between the well-known Eq.~(\ref{eq:Cs}), for HBT, and Eq.~(\ref{eq:B2}), for the D coalescence factor. This relation was derived first in~\cite{Blum:2019suo} starting from the QM limit. Our independent derivation here gives another perspective on this result, showing, for example, that it does not require density matrix factorisation to apply for its validity. 
In Sec.~\ref{ss:factorisation} we review how adding the assumption of density matrix factorisation allows one to connect two-particle HBT analyses to single-particle spectra and three-body coalescence. 
In Sec.~\ref{sss:HTHe3} we derive the coalescence prediction for $_\Lambda^3$H and $^3$He. Eqs.~(\ref{eq:B3H},\ref{eq:B3}), or their momentum space versions of Eqs.~(\ref{eq:B3Hmom},\ref{eq:B3mom}), are the most model-independent versions of these coalescence factors we know of.

None of our results in Sec.~\ref{s:basics} rely on the details of the underlying nucleon emission function, or needs to specify a model of the dynamical evolution of the HXS: our results simply connect femtoscopy with cluster yields, and the connection should apply to any self-consistent HXS model. To make contact with measurements, however, we need to specify the two- and three-particle source. 
We turn to this in Sec.~\ref{s:hbtB}. We start in Sec.~\ref{ss:GS} by appealing to observational HBT parameterisations of the two-particle source (extended, with some added assumptions, to the three-particle source). Relying on experimental fits allows us to keep the analysis as model-independent as we can. While this is not an essential requirement of the framework, we stick in this paper to Gaussian or semi-Gaussian parameterisations. In Apps.~\ref{ss:SnonG} and~\ref{sss:model} we give some quantitative model-dependent theoretical examples that suggest that an anisotropic (3D) Gaussian source parameterisation is probably accurate enough for the purpose of testing the origin of clusters via the coalescence--correlations relation.

The next ingredient needed is the nuclear wave functions. In Sec.~\ref{ss:waveGaus} we derive the coalescence factors corresponding to the simplified Gaussian wave functions. The Gaussian wave function is an over-simplification in some cases, but at the cost of an $\mathcal{O}(1)$ theoretical error it allows us to derive analytic results for the coalescence factors, summarised by Eqs.~(\ref{eq:B2G},\ref{eq:B3G},\ref{eq:B3HG}). We note that the Gaussian wave functions we consider allow for different cluster length scales for three-body states; for $_\Lambda^3$H this is crucial, as the p-n factor of the wave function is considerably more compact than the $\Lambda$-pn factor. Eqs.~(\ref{eq:B2G},\ref{eq:B3G},\ref{eq:B3HG}) also account for the intrinsically anisotropic shape of the two- and three-particle source describing the HXS. As we illustrate in App.~\ref{sss:model}, the two-particle source is expected to be truly anisotropic, especially at large $p_t$. 

In Sec.~\ref{ss:waveacc} we extend the analysis to more accurate non-Gaussian wave functions. For D we derive an analytic coalescence factor formula that applies to the Hulthen wave function if the underlying two-particle source is approximated as 1D Gaussian. The more realistic 3D two-particle source can be easily accounted for by numerical integration. For $_\Lambda^3$H we consider the recent three-body wave function proposed by~\cite{Hildenbrand:2019sgp}. We show that while the theoretically expected wave function is non-Gaussian, exhibiting an extended high-$q$ tail, nevertheless an effective Gaussian wave function fit does a reasonably accurate job in the coalescence factor calculation (valid to a factor of $\sim2$). What does turn out -- as already noted above -- to be quantitatively important, is the consideration of the two different length scales associated with the p-n and $\Lambda$-pn factors of the state.

In Sec.~\ref{s:dat} we give a rudimentary comparison to data.
In Sec.~\ref{ss:DHe3dat} we recap results from~\cite{Blum:2019suo} for D and $^3$He, adding a pPb data point to the PbPb and pp measurements discussed there; we also correct a few typos in~\cite{Blum:2019suo}. In Sec.~\ref{ss:HTdat} we compare the coalescence prediction for $^3_\Lambda$H to PbPb data. We emphasise (as was done before us~\cite{Zhang:2018euf,Sun:2018mqq,Bellini:2018epz}; but here with a robust coalescence calculation) that $^3_\Lambda$H data in small systems (pp, pPb, or low multiplicity PbPb) has the potential to conclusively rule out (or support) coalescence as the dominant origin of clusters.

In Sec.~\ref{s:conc} we discuss and summarise our results.

\section{The coalescence/femtoscopy framework}\label{s:basics}
The description of femtoscopic correlations between nucleons~\cite{Koonin:1977fh} and the coalescence model for nuclei~\cite{Bond:1977fd,Sato:1981ez} are two aspects of the same theoretical framework. The idea is that at kinetic freeze out the HXS can be described by a multi-particle density matrix $\hat\rho_{\rm HX}$. Depending on the measured observable, this density matrix can be projected onto final states of different multiplicities. 
In what follows we present a concise derivation of coalescence and the coalescence--correlations relation~\cite{Blum:2019suo}, aiming to collect different aspects of the problem under the same roof, so to speak. Many of the results were also derived elsewhere, notably in~\cite{Lednicky:1981su,Akkelin:2001nd,Lednicky:2005tb,Blum:2019suo} and  (albeit with model-dependence) in~\cite{Mrowczynski:1987oid,Mrowczynski:1989jd,Mrowczynski:1992gc,Mrowczynski:1993cx,Mrowczynski:1994rn,Maj:2004tb,Mrowczynski:2016xqm,Scheibl:1998tk,Blum:2017qnn}.

\subsection{Two-particle correlations and the deuteron.}\label{ss:2ptD}

If the phase space density of nucleons around the time when they last scatter against other particles (mostly pions) in the HXS is not too high, then the subsequent propagation of pairs of nucleons emitted very near in phase space would be dominated by FSI, while additional interactions with particles other than the pair would be sub-dominant. 
With this {\it sudden approximation}\footnote{See~\cite{Bond:1977fd,Pratt:1987zz,Csernai:1986qf} for nonrelativistic formulations. See~\cite{Martin:1996mv} for a discussion of corrections to the sudden approximation due to the residual charge of the HXS, as applied to low energy (80~MeV/nuc) HIC.}, the Lorentz-invariant yield of nucleon pairs at total spin $s$ is given by~\cite{Lednicky:1981su,Lednicky:2005tb}
\be\label{eq:2pLI} \gamma_1\gamma_2\frac{dN_{2,s}}{d^3{\bf p}_1d^3{\bf p}_2}&=&\frac{2s+1}{(2\pi)^6}\int d^4x_1\int d^4x_2\int d^4x'_1\int d^4x'_2\,\Psi^{*}_{s,p_1,p_2}(x'_1,x'_2)\,\Psi_{s,p_1,p_2}(x_1,x_2)\,\rho_{p_1,p_2}\left(x_1,x_2;x'_1,x'_2\right),
\ee
where $\Psi_{s,p_1,p_2}(x_1,x_2)$ is the continuum Bethe-Salpeter amplitude~\cite{Schweber:1955zz} describing the FSI of the pair. 
Similarly, the yield of deuterons at momentum $P$ is given by
\be\label{eq:DLI} \gamma\frac{dN_{\rm d}}{d^3{\bf P}}&=&\frac{2s_d+1}{(2\pi)^3}\int d^4x_1\int d^4x_2\int d^4x'_1\int d^4x'_2\,\Psi^{*}_{ d,P}(x'_1,x'_2)\,\Psi_{d,P}(x_1,x_2)\,\rho_{p_1,p_2}\left(x_1,x_2;x'_1,x'_2\right),
\ee
where $\Psi_{ d,P}(x_1,x_2)$ is the bound state Bethe-Salpeter amplitude describing the deuteron. 
The role of the Bethe-Salpeter amplitudes $\Psi$ is to resum soft diagonal (ladder) FSI diagrams, factoring their effect out of an assumed underlying short-distance amplitude forming the density matrix $\rho_{p_1,p_2}$. 
 
%

Obviously, modelling the spectra Eqs.~(\ref{eq:2pLI}-\ref{eq:DLI}) requires proper modelling of the FSI, that are either calculable from first principles (in the case of Coulomb) or measurable (in the case of nuclear scattering amplitudes) and that we assume to be known. Modelling $\rho_{p_1,p_2}$ from first principles, however, is currently impossible. Our goal in this section, and in the rest of the paper, is to demonstrate that even without a-priori knowledge of $\rho_{p_1,p_2}$, the mere fact that the same $\rho_{p_1,p_2}$ occurs in both of Eqs.~(\ref{eq:2pLI}) and~(\ref{eq:DLI}) is enough to allow a model-independent, approximate prediction of the deuteron (and, with some added assumptions, other clusters) spectrum based on measurements of HBT correlations~\cite{Blum:2019suo}.

Let us define $c_{1,2}=(p_{1,2} P)/P^2$, where the pair total momentum is $P=p_1+p_2\,\equiv \,2p$. In general, the dependence of the amplitude $\Psi$ (be it $\Psi_{s,p_1,p_2}$ or $\Psi_{ d,P}$) on the pair total momentum and centre of mass coordinate $X=c_1x_1+c_2x_2$ 
can be factored out from the dependence on the relative momentum\footnote{Note $c_1+c_2=1$, $qP=0$ and $(p_1-p_2)P=(c_1-c_2)P^2=m_1^2-m_2^2$.} $q=c_2p_1-c_1p_2$ and relative position $x=x_1-x_2$, via 
\be \label{eq:Psifact}\Psi(x_1,x_2)&=&e^{-iPX}\phi(x).\ee
Using Eq.~(\ref{eq:Psifact}) 
and changing to convenient coordinates, we can rewrite Eqs.~(\ref{eq:2pLI}-\ref{eq:DLI}) as:
\be\label{eq:2pLIDk} \gamma_1\gamma_2\frac{dN_{2,s}}{d^3{\bf p}_1d^3{\bf p}_2}&=&\frac{2s+1}{(2\pi)^6}\int d^4{ r}\int \frac{d^4{ k}}{(2\pi)^4}\,\tilde{\mathcal{D}}_{s, q}\left({ k},{ r}\right)\,\tilde S_{p_1,p_2}\left({ k,r}\right),\\
\label{eq:DLIDk}\gamma\frac{dN_{\rm d}}{d^3{\bf P}}&=&\frac{2s_d+1}{(2\pi)^3}\int d^4{ r}\int \frac{d^4{ k}}{(2\pi)^4}\,\tilde{\mathcal{D}}_{ d}\left({ k},{ r}\right)\,\tilde S_{p_1,p_2}\left({ k,r}\right),
\ee
where we define the relativistic internal Wigner density 
\be\label{eq:D}\tilde{\mathcal{D}}\left( { k}, { r}\right)&=&\int d^4{\zeta}\,e^{i { k}{\bf\zeta}}\,\phi\left( { r}+\frac{{\zeta}}{2}\right)\,\phi^*\left( { r}-\frac{{\zeta}}{2}\right)
\ee
(with $\tilde{\mathcal{D}}_{s, q}$ and $\tilde{\mathcal{D}}_{ d}$ obtained from $\phi_{s,q}$ and $\phi_{ d}$, respectively) and where 
\be\label{eq:Stilde} \tilde S_{p_1,p_2}\left({ k,r}\right)&=&\int d^4x\int d^4l_1\,e^{-il_1(c_1P+k)}\int d^4l_2\,e^{-il_2(c_2P-k)}\,\times\\
&&\rho_{p_1,p_2}\left(x+c_2r+\frac{l_1}{2},x-c_1r+\frac{l_2}{2};x+c_2r-\frac{l_1}{2},x-c_1r-\frac{l_2}{2}\right).\no
\ee

The Wigner density we would obtain if we could turn off both FSI and quantum statistics is $\D_{s,q}^{0}(k,r)=(2\pi)^4\delta^{(4)}\left(k-q\right)$, independent of $r$. With this we can define a hypothetical reference uncorrelated pair spectrum,
\be\label{eq:2pLIDk0} \gamma_1\gamma_2\frac{dN_{2}^0}{d^3{\bf p}_1d^3{\bf p}_2}&=&\frac{(2s_N+1)^2}{(2\pi)^6}\int d^4{ r}\,\tilde S_{p_1,p_2}\left(q,r\right),
\ee
where $s_N=1/2$ is the nucleon spin. The reference pair spectrum is not a real physical object, but we can try to mimic it experimentally by pairing particles from different events with similar event characteristics.
With this understanding, the pair correlation function is defined as
\be\label{eq:C0}
C(p,q)&=&\frac{\sum_s\gamma_1\gamma_2\frac{dN_{2,s}}{d^3{\bf p}_1d^3{\bf p}_2}}{\gamma_1\gamma_2\frac{dN_{2}^0}{d^3{\bf p}_1d^3{\bf p}_2}}
\ee
and the coalescence factor for deuteron formation is defined as
\be\label{eq:B20}\B_2(p)&=&\frac{P^0\frac{dN_{\rm d}}{d^3{\bf P}}}{p^0_1p^0_2\frac{dN_{2}^0}{d^3{\bf p}_1d^3{\bf p}_2}}
\;\approx\;\frac{2}{m}\frac{\gamma\frac{dN_{\rm d}}{d^3{\bf P}}}{\gamma_1\gamma_2\frac{dN_{2}^0}{d^3{\bf p}_1d^3{\bf p}_2}},\ee
where we approximated $m_D\approx2m$.

At this point it is useful to make two approximations:
\begin{itemize}
\item {\bf Smoothness approximation:}  The smoothness approximation was discussed widely in the literature~\cite{Pratt:1997pw,Heinz:1999rw,Lisa:2005dd}. The version of this approximation we take here amounts to replacing $S_{p_1,p_2}(k,r)\approx S_{p_1,p_2}(0,r)$ in Eqs.~(\ref{eq:2pLIDk}) and~(\ref{eq:DLIDk}). The $k$ integral of the $\tilde{\D}$ functions can then be done, yielding $\int \frac{d^4k}{(2\pi)^4}\tilde{\D}(k,r)=|\phi(r)|^2$. Similarly, we replace $S_{p_1,p_2}(q,r)\approx S_{p_1,p_2}(0,r)$ in the reference pair spectrum Eq.~(\ref{eq:2pLIDk0}).
The accuracy of the smoothness approximation is probably sufficient for our purpose in PbPb collisions; in pp collisions we think that a careful assessment is still warranted.

\item {\bf Equal-time approximation:} In the pair rest frame (PRF) we have $P=(M,{\bf 0})$, $q=(0,{\bf q})$ and $x=(t,{\bf x})$. 
Both for two-particle correlations and bound states, we are interested in pairs that are nonrelativistic in the PRF, ${\bf q}^{2}\ll m^2$, and can neglect corrections of $\mathcal{O}({\bf q}^2/m^2)$. 
A key point, derived clearly in~\cite{Lednicky:1981su,Lednicky:2005tb}, is that the Bethe-Salpeter amplitude in the PRF-nonrelativistic limit is approximately independent of the PRF time:
\be\label{eq:sudden}\phi(x)&=&\phi({\bf x})\left(1+\mathcal{O}\left(\frac{t}{m{\bf x}^{2}}\right)\right).\ee
At the level of the leading term in the equal-time approximation of Eq.~(\ref{eq:sudden}), $\phi_{s,q}$ ($\phi_{ d}$) is equal to the QM static scattering wave (bound state) solution of the Schr\"odinger equation~\cite{Landau:1991wop}. 

In hadronic collisions we have $t\sim {\bf x}\sim$~few fm, implying a correction $\sim0.2\frac{1\rm fm}{{\bf x}}$ to the equal-time approximation. Thus, we are certainly sacrificing some precision when we adopt it: the incurred theoretical error is probably in the ballpark of 10\% for PbPb collisions, where ${\bf x}\sim3$~fm, but could be several tens of percent in pp collisions where ${\bf x}\sim1$~fm. Nevertheless, the equal-time approximation would allow us significant mileage with relatively simple notation. We think that it probably allows for sufficient accuracy to establish (or exclude) coalescence as the main origin of clusters in hadronic collisions. Note that the free solution in the absence of FSI is the plane wave $\phi(x)=e^{-iqx}$ [(anti-)symmetrised by QS] and the equal-time limit of Eq.~(\ref{eq:sudden}) is exact in that case, because in the PRF $qx=-{\bf qx}$.
\end{itemize}

Adopting the smoothness and equal-time approximations we are led to the definition\footnote{It is common practice (e.g.~\cite{Adam:2015vja,Acharya:2018gyz,Acharya:2020dfb}) to drop the explicit mention of ${ p}$ in $\mathcal{S}_2({\bf r})$; this, despite the fact that $\mathcal{S}_2({\bf r})$ does depend on ${ p}$. As long as we keep this fact in mind, this practice brings no harm.} of the normalised two-particle source $\mathcal{S}_2$, a function of the PRF spatial coordinate ${\bf r}$, as an integral in PRF time $t=r^0$:
\be \mathcal{S}_2({\bf r})&=&\frac{\int dr^0\tilde S_{p_1,p_2}\left({0,r}\right)}{\int d^4{ r}\,\tilde S_{p_1,p_2}\left({0,r}\right)}.\ee
The pair correlation function is then given from Eq.~(\ref{eq:C0}) as
\be\label{eq:C} C(p,q)&=&\sum_sw_sC_s(p,q)
\ee
with the spin weights $w_s=(2s+1)/(2s_N+1)^2$ and
\be\label{eq:Cs}
C_s(p,q)&\approx&\int d^3{\bf r}\,\left|\phi_{s, q}\left({\bf r}\right)\right|^2\mathcal{S}_2({\bf r}).\ee
Similarly, the deuteron coalescence factor is given from Eq.~(\ref{eq:B20}) as
\be\label{eq:B2}\B_2(p)
&\approx&\frac{2(2s_d+1)}{m(2s_N+1)^2}(2\pi)^3\int d^3{\bf r}\,\left|\phi_{d}\left({\bf r}\right)\right|^2\,\mathcal{S}_2({\bf r}).\ee

Two general comments are in order. 
First, the aim of HBT analyses like, e.g., Refs.~\cite{Adam:2015vja,Acharya:2018gyz,Acharya:2020dfb} (for reviews, see~\cite{Lisa:2005dd,Heinz:1999rw}) is to measure the source $\mathcal{S}_2({\bf r})$ by solving the Schr\"odinger equation for the wave functions $\phi_{s, q}({\bf r})$~\cite{Koonin:1977fh,Lednicky:2005tb,Mihaylov:2018rva} and comparing the observed two-particle spectrum with Eqs.~(\ref{eq:C}-\ref{eq:Cs}). In this exercise, obtaining agreement with the experimental data appears to require the use of the correct set of FSI potentials. While this point is not often stated explicitly, the success of the source reconstruction analyses lends some credence to the basic framework leading to Eqs.~(\ref{eq:C}-\ref{eq:Cs}). As we have just seen, this is the same framework that stands behind the coalescence model for nuclei, the only difference being that HBT deals with continuum scattering state solutions of FSI and coalescence deals with discrete bound state solutions of FSI. Thus the mere existence of successful HBT analyses lends some credence to coalescence as the origin of (at least some of the) nuclei observed in hadronic collisions.

Second, with $\mathcal{S}_2({\bf r})$ measured, Eq.~(\ref{eq:B2}) predicts the deuteron yield model-independently and with no free parameters. As we show in Sec.~\ref{s:dat}, this prediction is consistent with experimental data to within a factor of two or so across systems ranging from PbPb to pPb and pp at different multiplicities~\cite{Blum:2017qnn,Blum:2019suo}. This is strong evidence that coalescence contributes to the production of deuterons in hadronic collisions at the $\mathcal{O}(1)$ level, at least.

A caveat to keep in mind is that although Eq.~(\ref{eq:Cs}) is commonly used in the literature, we do not know a model-independent way of checking the quantitative corrections due to the smoothness and equal-time approximations, which we have done to reduce Eq.~(\ref{eq:2pLIDk}) into Eq.~(\ref{eq:Cs}); nor to the sudden approximation itself, allowing us to write Eq.~(\ref{eq:2pLIDk}) in the first place. A systematic study of these uncertainties is warranted if one wishes to narrow down the theory uncertainty associated with Eq.~(\ref{eq:B2}).

As a technical aside, it is convenient to carry out some of the analysis in momentum space. To this end it is useful to introduce the momentum space correlation function $\mathcal{C}_2$, which is just the Fourier transform of $\mathcal{S}_2({\bf r})$:
\be\label{eq:C2mom} \mathcal{C}_2(p, {\bf k})&=&\int d^3{\bf r}\,e^{i{\bf kr}}\,\mathcal{S}_2({\bf r}).\ee
(We will usually keep the explicit appearance of $p$ in $\mathcal{C}_2$.) 
Defining the momentum space deuteron form factor $\F_d$
\be\label{eq:D(k)} \left|\phi_d\left( {\bf r}\right)\right|^2&=&\int \frac{d^3{\bf k}}{(2\pi)^3}\,e^{i {\bf k} {\bf r}}\,\mathcal{F}_d\left( {\bf k}\right)\ee
we can rewrite Eq.~(\ref{eq:B2}) as
\be\label{eq:B2mom}\B_2(p)
&\approx&\frac{2(2s_d+1)}{m(2s_N+1)^2}\int d^3{\bf k}\,\mathcal{F}_d\left( {\bf k}\right)\mathcal{C}_2\left(p,{\bf k}\right).\ee
This is the version of the coalescence--correlation relation derived in Ref.~\cite{Blum:2019suo}.

\subsection{Connecting two-particle states with one- and three-particle states: density matrix factorisation.}\label{ss:factorisation}

While two-particle HBT correlations are directly and model-independently connected to deuteron coalescence, the connection to single-particle spectra and to the coalescence of three-body states requires further assumptions. The main assumption we need is the factorisation of the multi-particle density matrix into the product of single-particle density matrices, e.g.~\cite{Koonin:1977fh,Sato:1981ez}
\be\label{eq:rhofact}\rho_{p_1,p_2}\left(x_1,x_2;x_1',x_2'\right)&\approx&\rho_{p_1}(x_1;x_1')\rho_{p_2}(x_2;x_2').\ee
Factorisation cannot be exact. For example, in low multiplicity events total momentum conservation inevitably leads to the breakdown of Eq.~(\ref{eq:rhofact}), seen observationally as ``non-femtoscopic correlations" at large ${\bf q}$. Nevertheless, keeping these caveats in mind we will adopt the factorisation approximation in what follows. Later on we investigate one simple way to parametrise related corrections.

\subsubsection{Uncorrelated pair spectrum as a product of one-particle spectra.}
Considering nucleons (protons and neutrons) with $m_1\approx m_2\approx m$, nonrelativistic in the PRF, we have $q\approx p_1-p_2$ and the factors $c_1\approx c_2=\frac{1}{2}+\mathcal{O}\left({\bf q}^2/m^2\right)$. Inserting this into Eq.~(\ref{eq:Stilde}) and using Eq.~(\ref{eq:rhofact}) leads to
\be \tilde S_{p_1,p_2}\left(q,r\right)&\approx&\int d^4x\,\tilde S_{p_1}\left(x+\frac{r}{2}\right)\tilde S_{p_2}\left(x-\frac{r}{2}\right),
\ee
with the one-particle emission function
\be \tilde S_p(x)&=&\int d^4l\,e^{-ilp}\rho_{p}\left(x+\frac{l}{2};x-\frac{l}{2}\right).\ee
This $\tilde S_p(x)$ coincides (up to constant factors in the definition) with the particle source of~\cite{Koonin:1977fh} and with the emission function or phase space density of~\cite{Anchishkin:1997tb,Wiedemann:1997cn,Scheibl:1998tk,Lisa:2005dd,Heinz:1999rw}. 
The reference pair spectrum factorises into the product of single-particle spectra,
\be\label{eq:2pLIDk0fct} \gamma_1\gamma_2\frac{dN_{2}^0}{d^3{\bf p}_1d^3{\bf p}_2}&\approx&\left[\gamma_1\frac{dN}{d^3{\bf p}_1}\right]\left[\gamma_2\frac{dN}{d^3{\bf p}_2}\right],
\ee
\be\label{eq:1p}
\gamma\frac{dN}{d^3{\bf p}}&=&\frac{(2s_N+1)}{(2\pi)^3}\int d^4x\,\tilde S_p(x).
\ee

Finally, the two-particle source $\mathcal{S}_2$ is constructed from single-particle emission functions as
\be\label{eq:S2fact} \mathcal{S}_2({\bf r})&=&\frac{\int dr^0\int d^4x\,\tilde S_{p}\left(x+\frac{r}{2}\right)\tilde S_{p}\left(x-\frac{r}{2}\right)}{\left[\int d^4x\,\tilde S_p(x)\right]^2}.\ee
It can be more convenient to calculate $\C_2(p,{\bf q})$, by inserting Eq.~(\ref{eq:S2fact}) into Eq.~(\ref{eq:C2mom}), giving the prescription:
\be \label{eq:C2model}\C_2(p,{\bf q})&=&\frac{\left|\int d^4x\,e^{iqx}\tilde S_{p}\left(x\right)\right|^2}{\left[\int d^4x\,\tilde S_p(x)\right]^2}.\ee
In evaluating Eq.~(\ref{eq:C2model}), recall that we require $q=(0,{\bf q})$ as specified in the PRF.  

As a slight detour, consider the proton pair correlation with FSI turned off but quantum statistics still on, in the spin-asymmetric or spin-symmetric state where $\phi_{s,q}({\bf r})=\frac{1}{\sqrt{2}}\left(e^{i{\bf qr}}\pm e^{-i{\bf qr}}\right)$, respectively. Using Eq.~(\ref{eq:S2fact}) and noting that $q^0=0$ in the PRF, the pair correlation of Eq.~(\ref{eq:Cs}) would be
\be C_s(p,q)&\approx&\int d^3{\bf r}\left|\phi_{s,q}({\bf r})\right|^2\mathcal{S}_2({\bf r})\;=\;1\pm\frac{\left|\int d^4x\,e^{2iqx}\tilde S_{p}\left(x\right)\right|^2}{\left[\int d^4x\,\tilde S_p(x)\right]^2},\ee
consistent with the usual expression in the literature~\cite{Chapman:1995nz,Wiedemann:1997cn,Heinz:1999rw,Lisa:2005dd} (note that $q$ as defined in, e.g.~\cite{Chapman:1995nz} is equal to $2q$ in our notation).

\subsubsection{Hypertriton and $^3$He.}\label{sss:HTHe3}

The starting point in the coalescence calculation for hypertriton ${\rm ^3_\Lambda H}$ (pn$\Lambda$) is similar to Eq.~(\ref{eq:DLI}) for the deuteron:
\be\label{eq:HLI} \gamma\frac{dN_{{\rm ^3_\Lambda H}}}{d^3{\bf P}}&=&\frac{2s_{\rm ^3_\Lambda H}+1}{(2\pi)^3}\int d^4x_p\int d^4x_n\int d^4x_\Lambda\int d^4x'_p\int d^4x'_n\int d^4x'_\Lambda\,\times\no\\
&&\Psi^{*}_{{\rm ^3_\Lambda H},P}(x'_p,x'_n,x'_\Lambda)\,\Psi_{{\rm ^3_\Lambda H},P}(x_p,x_n,x_\Lambda)\,\rho_{p_p,p_n,p_\Lambda}\left(x_p,x_n,x_\Lambda;x'_p,x'_n,x'_\Lambda\right),
\ee
where $\Psi_{{\rm ^3_\Lambda H},P}(x_p,x_n,x_\Lambda)$ is the bound state Bethe-Salpeter amplitude describing the ${\rm ^3_\Lambda H}$. 
The total momentum is $P=p_p+p_n+p_\Lambda\equiv3p$. 
We also define $P_{pn}=p_p+p_n$, 
$c_I=(p_IP)/P^2$ with $I=p,n,\Lambda$, and $\tilde c_J=(p_JP_{pn})/P_{pn}^2$ with $J=p,n$. 
The centre of mass coordinate is then $X=c_nx_n+c_px_p+c_\Lambda x_\Lambda$, and useful relative coordinates are $r_{pn}=x_p-x_n$ and $r_\Lambda=x_\Lambda-\tilde c_px_p-\tilde c_nx_n$. With these definitions the Bethe-Salpeter amplitude factorises into $\Psi_{{\rm ^3_\Lambda H},P}=e^{-iPX}\phi_{{\rm ^3_\Lambda H}}(r_{pn},r_\Lambda)$. 

The calculation for $^3$He (ppn) is similar, with the replacements $r_{pn}\to r_{pp}$ and $r_\Lambda\to r_n$, etc.

Following the same steps as in Sec.~\ref{ss:2ptD} and adding to that the smoothness approximation, the equal-time approximation, as well as density matrix factorisation a-la Eq.~(\ref{eq:rhofact}) extended to three particles, we are led -- after the dust settles -- to the normalised three-particle source expressed as integrals of single-particle emission functions,
\be\label{eq:S3H} \mathcal{S}_{3\Lambda}({\bf r}_{pn},{\bf r}_\Lambda)&=&\frac{\int dr_{pn}^0\int dr_{\Lambda}^0\int d^4x\,\tilde S_{p}\left(x+\frac{{ r}_{pn}}{2}-\frac{ { r}_\Lambda}{3}\right)\tilde S_{p}\left(x-\frac{{ r}_{pn}}{2}-\frac{ { r}_\Lambda}{3}\right)\tilde S^{(\Lambda)}_{p}\left(x+\frac{2r_\Lambda}{3}\right)}{\left[\int d^4x\,\tilde S_p(x)\right]^2\int d^4x\,\tilde S^{(\Lambda)}_{p}(x)},\ee
\be\label{eq:S3} \mathcal{S}_{3}({\bf r}_{pp},{\bf r}_n)&=&\frac{\int dr_{pp}^0\int dr_{n}^0\int d^4x\,\tilde S_{p}\left(x+\frac{{ r}_{pp}}{2}-\frac{ { r}_n}{3}\right)\tilde S_{p}\left(x-\frac{{ r}_{pp}}{2}-\frac{ { r}_n}{3}\right)\tilde S_{p}\left(x+\frac{2r_n}{3}\right)}{\left[\int d^4x\,\tilde S_p(x)\right]^3}.\ee
The coalescence factors are then found as
\be\label{eq:B3H}
\B_{3\Lambda}(p)&\approx&\frac{3}{m^2}\frac{2s_{\rm ^3_\Lambda H}+1}{(2s_N+1)^3}(2\pi)^6\int d^3{\bf r}_{pn}\int d^3{\bf r}_{\Lambda}\,\left|\phi_{\rm ^3_\Lambda H}\left({\bf r}_{pn},{\bf r}_\Lambda\right)\right|^2\,\mathcal{S}_{3\Lambda}\left({\bf r}_{pn},{\bf r}_\Lambda\right),
\ee
\be\label{eq:B3}
\B_{3}(p)&\approx&\frac{3}{m^2}\frac{2s_{\rm He}+1}{(2s_N+1)^3}(2\pi)^6\int d^3{\bf r}_{pp}\int d^3{\bf r}_{n}\,\left|\phi_{\rm ^3He}\left({\bf r}_{pp},{\bf r}_{n,}\right)\right|^2\,\mathcal{S}_3\left({\bf r}_{pp},{\bf r}_n\right).
\ee
We highlight that in the $\mathcal{S}_{3\Lambda}$ calculation one emission function corresponds to the emission of a $\Lambda$, rather than a nucleon. For simplicity we approximated $m_\Lambda\approx m$ in the prefactor, at the cost of an error of about $\sim10\%$.

Again, we can define the Fourier transform
\be
\C_{3\Lambda}\left(p,{\bf q}_1,{\bf q}_2\right)&=&\int d^3{\bf r}_1\int d^3{\bf r}_2\,e^{i {\bf q}_1 {\bf r}_1+i {\bf q}_2 {\bf r}_2}\mathcal{S}_{3\Lambda}({\bf r}_1,{\bf r}_2)
\ee
and the momentum space form factor $\F_{\rm ^3_\Lambda H}$,
\be\left|\phi_{{\rm ^3_\Lambda H}}\left({\bf r}_{pn},{\bf r}_{\Lambda}\right)\right|^2&=&\int\frac{d^3{\bf k}_{pn}}{(2\pi)^3}e^{i{\bf k}_d{\bf r}_{pn}}\int\frac{d^3{\bf k}_\Lambda}{(2\pi)^3}e^{i{\bf k}_\Lambda {\bf r}_\Lambda}\F_{\rm ^3_\Lambda H}({\bf k}_{pn},{\bf k}_\Lambda).\ee
In terms of these (and their equivalents for $^3$He) the coalescence factors read
\be\label{eq:B3Hmom}
\B_{3\Lambda}(p)&\approx&\frac{3}{m^2}\frac{2s_{\rm ^3_\Lambda H}+1}{(2s_N+1)^3}\int d^3{\bf k}_{pn}\int d^3{\bf k}_{\Lambda}\,\F_{\rm ^3_\Lambda H}({\bf k}_{pn},{\bf k}_\Lambda)\,\C_{3\Lambda}\left(p,{\bf k}_{pn},{\bf k}_\Lambda\right),\\
\label{eq:B3mom}
\B_{3}(p)&\approx&\frac{3}{m^2}\frac{2s_{\rm ^3He}+1}{(2s_N+1)^3}\int d^3{\bf k}_{pp}\int d^3{\bf k}_{n}\,\F_{\rm ^3He}({\bf k}_{pp},{\bf k}_n)\,\C_{3}\left(p,{\bf k}_{pp},{\bf k}_n\right).
\ee

The coalescence of hypertriton and $^3$He can be connected to HBT analyses, to the extent that the three-particle normalised sources $\mathcal{S}_{3\Lambda}$ and $\mathcal{S}_{3}$ can be accessed by correlation measurements. This connection would be crucial for the attempt to use ${\rm ^3_\Lambda H}$, due to its large nucleus size, as a test of the coalescence framework. However, the connection is not as direct as it is for deuteron formation. We discuss this connection further in the next section.

\section{Coalescence from correlation functions}\label{s:hbtB}

\subsection{HBT source parameterisation.}\label{ss:GS}
We now want to make practical contact with observational information on the HXS source size, available from HBT studies. Experimental analyses commonly fit the measured two-particle source using a Gaussian approximation~\cite{Adam:2015vja,Acharya:2018gyz,Acharya:2020dfb}, the simplest version of which is the isotropic (or 1D) Gaussian~\cite{Lednicky:1981su}
\be\label{eq:S_2(r)isoGauss} \mathcal{S}^{1D}_2({\bf r})&=&\frac{1}{\left(4\pi R_{\rm inv}^2\right)^{\frac{3}{2}}}e^{-\frac{{\bf r}^2}{4R_{\rm inv}^2}}.
\ee
The radius parameter $R_{\rm inv}$ depends on ${ p}$~\cite{Adam:2015vja,Acharya:2018gyz,Acharya:2020dfb}.

From the theoretical perspective~\cite{Pratt:1997pw,Heinz:1999rw,Lisa:2005dd} Eq.~(\ref{eq:S_2(r)isoGauss}) could arise, for example, if the emission function $\tilde S_p(x)$ receives most of its support near some configuration space location ${\bf R}_s$ and is also sharply peaked around kinetic freeze out time $t_f$, as measured in the emitted particle rest frame, allowing the approximation
\be\label{eq:fGauss} \tilde S_p\left(x\right)&\propto&e^{-\frac{(\bf r-{\bf R}_s)^2}{2R_{\rm inv}^2}}\delta(t-t_f).\ee
Inserting Eq.~(\ref{eq:fGauss}) into Eq.~(\ref{eq:S2fact}) immediately yields Eq.~(\ref{eq:S_2(r)isoGauss}). 
While more difficult to test experimentally~\cite{Abelev:2014pja}, Eq.~(\ref{eq:fGauss}) also predicts $\mathcal{S}_3$ of Eqs.~(\ref{eq:S3H}) and~(\ref{eq:S3}),
\be\label{eq:S3ss} \mathcal{S}^{1D}_{3}({\bf r}_1,{\bf r}_2)&=&\frac{\int d^3{\bf x}\,e^{-\frac{\left({\bf x}+\frac{{\bf r}_{1}}{2}-\frac{ {\bf r}_2}{3}-{\bf R}_s\right)^2+\left({\bf x}-\frac{{\bf r}_{1}}{2}-\frac{ {\bf r}_2}{3}-{\bf R}_s\right)^2+\left({\bf x}+\frac{2{\bf r}_2}{3}-{\bf R}_s\right)^2}{2R_{\rm inv}^2}}}{\left[\int d^3{\bf x}\,e^{-\frac{(\bf x-{\bf R}_s)^2}{2R_{\rm inv}^2}}\right]^3}
\;\;=\;\;\frac{1}{\left(12\pi^2 R_{\rm inv}^4\right)^{\frac{3}{2}}}e^{-\frac{{\bf r}_1^2+\frac{4}{3}{\bf r}_2^2}{4R_{\rm inv}^2}}.\ee
The momentum space versions of the isotropic Gaussian source model are
\be\label{eq:C2ss} \C^{1D}_2(p,{\bf q})&=&e^{-{\bf q}^2R_{\rm inv}^2},\\
\label{eq:C3ss}\C^{1D}_3\left(p,{\bf q}_1,{\bf q}_2\right)&=&e^{-R_{\rm inv}^2\left({\bf q}_1^2+\frac{3}{4}{\bf q}_2^2\right)}.
\ee

Note that here we have considered the correlation of particles (e.g. protons) with the same underlying emission function $\tilde S_p(x)$. Soon, however, we will use Eq.~(\ref{eq:C3ss}) to analyse ${\rm ^3_\Lambda H}$ production which involves both $\tilde S_p(x)$ and $\tilde S^{(\Lambda)}_p(x)$, and there is no guarantee that these emission functions involve the same values of ${\bf R}_s$ and $R_{\rm inv}$ in Eq.~(\ref{eq:fGauss}). 
Experimental results in pp collisions~\cite{Acharya:2020dfb} suggest that $R_{\rm inv}$ as extracted from p$\Lambda$ correlations actually differs from $R_{\rm inv}$ extracted from pp correlations by $\sim20\%$ or so\footnote{This is attributed in~\cite{Acharya:2020dfb} to the different contributions of strong resonance decays to the p and $\Lambda$ spectra. From the point of view of our discussion, however, the cause of the difference in $R_{\rm inv}$ is not essential.}. It must then be understood that any effective value we use for $R_{\rm inv}$ in Eqs.~(\ref{eq:S3ss}) or~(\ref{eq:C3ss}) cannot be more accurately determined than the aforementioned $\sim20\%$, without introducing model-dependent assumptions concerning the behaviour of the emission functions. We will comment on this point again in Sec.~\ref{ss:HTdat} when we discuss the comparison between the $^3_\Lambda$H coalescence prediction and experimental data. 

The isotropic Gaussian source model is quite unrealistic. Even if the HXS ``fire ball" was somehow isotropic in the lab frame (which it generally isn't; e.g.~\cite{Chapman:1995nz,Lisa:2005dd}), it would be seen as anisotropic in the PRF due to Lorentz contraction along the direction of ${\bf p}$. 
In addition, the beam line, of course, is a special direction in the initial state forming the HXS. 
As a simple generalisation that can capture some of these effects (and others), one can consider an anisotropic (3D) Gaussian source with a free normalisation\footnote{Adding off-diagonal components like $q_oq_lR^2_{ol}$, for example, into our analysis would be straightforward, but we avoid it here for simplicity.}:
\be\label{eq:C2ssng} \C^{3D}_2(p,{\bf q})&=&\lambda_2\,e^{-q_l^2R_l^2-q_o^2R_o^2-q_s^2R_s^2},\\
\label{eq:C3ssng}\C^{3D}_3\left(p,{\bf q}_1,{\bf q}_2\right)&=&\lambda_3\,e^{-R_l^2\left(q_{1l}^2+\frac{3}{4}q_{2l}^2\right)-R_o^2\left(q_{1o}^2+\frac{3}{4}q_{2o}^2\right)-R_s^2\left(q_{1s}^2+\frac{3}{4}q_{2s}^2\right)}.
\ee
Here, we split the 3-vector ${\bf q}$ into its three components: $q_o$ (``out") along the direction of the mean transverse momentum ${\bf p}_t$; $q_l$ (``longitudinal") along the beam axis; and $q_s$ (``side") along the third orthogonal direction ${\bf p}_t\times\hat z$.

Like the HBT radii $R_{o,s,l}$, the normalisation factors (``intercept" or chaoticity parameter~\cite{Wiedemann:1996ig,Akkelin:2001nd}) $\lambda_{2,3}$ should best be measured directly from the data. For two-nucleon correlations this is sometimes done~\cite{Adam:2015vja}. For three-nucleon correlations, as far as we know there is as yet no precedence. 

In App.~\ref{ss:SnonG} we briefly review how weak and strong resonance decays can distort the Gaussian shape of the source. In App.~\ref{sss:model} we give numerical examples of $\C_2$ calculated in the phenomenological blast wave model, illustrating how anisotropic flow and Lorentz contraction at $p_t>0$ give rise to a true 3D source. We find that while the 1D Eq.~(\ref{eq:C2ss}) can fail quite badly, the 3D Eq.~(\ref{eq:C2ssng}) as a phenomenological parameterisation is flexible enough to capture the true physical source to good accuracy.

Given the source parameterisation of Eqs.~(\ref{eq:C2ssng}-\ref{eq:C3ssng}), the final ingredient we need to evaluate Eqs.~(\ref{eq:B2mom}),~(\ref{eq:B3Hmom}) and~(\ref{eq:B3mom}) for the coalescence factors are the nucleus wave functions, encoded by the form factors $\mathcal{F}_{d,{\rm ^3He},{\rm ^3_\Lambda H}}$. We turn to that next.

\subsection{Nucleus wave functions: Gaussian wave function approximation.}\label{ss:waveGaus}
The results are particularly tractable if we make the simplifying approximation of Gaussian wave functions. This is useful for analytic insight and is also reasonable if one wants to test coalescence at the $\mathcal{O}(1)$ level, so we report the results in this section. In the next section we consider more accurate parameterisations of the wave functions.

For D we consider 
\be\phi_d({\bf r})&=&\left(\frac{1}{\pi b_d^2}\right)^{\frac{3}{4}}e^{-\frac{{\bf r}^2}{2b_d^2}},\ee
with momentum space form factor
\be\label{eq:FdG}\F_d({\bf k})&=&e^{-\frac{b_d^2{\bf k}^2}{4}}.\ee

For $^3$He we consider a Gaussian that is isotropic in the normalised Jacobi coordinates, related to our natural kinematic coordinates\footnote{One can verify that $\int d^3{\bf r}_{pp}\int d^3{\bf r}_{n}\left|\phi_{\rm ^3He}({\bf r}_{pp},{\bf r}_n)\right|^2=(3)^{\frac{3}{2}}\int d^3{\bf \eta}_{pp}\int d^3{\bf \eta}_{n}\left(\frac{1}{3\pi^2b_{\rm ^3He}^4}\right)^{\frac{3}{2}}e^{-\frac{{\bf \eta}_{pp}^2+{\bf\eta}_n^2}{b_{\rm ^3He}^2}}=1$.} via ${\bf \eta}_{pp}=\frac{1}{\sqrt{2}}{\bf r}_{pp}$, ${\bf \eta}_{n}=\sqrt{\frac{2}{3}}{\bf r}_{n}$:
\be\label{eq:phiHeJac}\phi_{\rm ^3He}({\bf r}_{pp},{\bf r}_n)&=&\left(\frac{1}{3\pi^2b_{\rm ^3He}^4}\right)^{\frac{3}{4}}e^{-\frac{{\bf r}_{pp}^2+\frac{4}{3}{\bf r}_n^2}{4b_{\rm ^3He}^2}}.\ee
With this $\phi_{\rm ^3He}$ the momentum space form factor is
\be\label{eq:FHe}\F_{\rm ^3He}({\bf k}_{pp},{\bf k}_n)&=&\int d^3{\bf r}_{pp}e^{-i{\bf k_{pp}r}_{pp}}\int d^3{\bf r}_ne^{-i{\bf k_nr_n}}\left|\phi_{\rm ^3He}\left({\bf r}_{pp},{\bf r}_{n}\right)\right|^2\;=\;e^{-\frac{b_{\rm ^3He}^2}{2}\left({\bf k}^2_{pp}+\frac{3}{4}{\bf k}_n^2\right)}.\ee

For ${\rm ^3_\Lambda H}$, as a first approximation we consider a product of Gaussians,
\be\label{eq:phiHTJac}\phi_{{\rm ^3_\Lambda H}}({\bf r}_{pn},{\bf r}_\Lambda)&\approx&\left(\frac{1}{3\pi^2b_{pn}^2b_\Lambda^2}\right)^{\frac{3}{4}}e^{-\frac{{\bf r}_{pn}^2}{4b_{pn}^2}-\frac{{\bf r}_\Lambda^2}{3b_\Lambda^2}}.\ee
The momentum space form factor is then
\be\label{eq:FHTGauss}\F_{\rm ^3_\Lambda H}({\bf k}_{pn},{\bf k}_\Lambda)&\approx&e^{-\frac{1}{2}\left(b_{pn}^2{\bf k}^2_{pn}+\frac{3}{4}b_\Lambda^2{\bf k}_\Lambda^2\right)}.\ee

We need to match the $b$ parameters to nuclear data. For D we take $b_d=3.5$~fm, corresponding to the RMS charge radius $r_{rms}=\sqrt{\frac{3}{8}}b_d=2.13$~fm~\cite{Sick:2015spa}. For an isotropic three-body Gaussian wave function, the parameter $b$ as we defined it is directly the RMS charge radius.   
For $^3$He this is $b_{\rm ^3He}\approx1.97$~fm~\cite{Sick:2015spa}. 
For ${\rm ^3_\Lambda H}$, Hildenbrand \& Hammer~\cite{Hildenbrand:2019sgp} reported\footnote{Please note that $\sqrt{\langle {\bf r}_{pn}^2\rangle}$ in~\cite{Hildenbrand:2019sgp} refers to the distance between the n and the p; that is, the diameter, not the radius of that subsystem. Namely, $\langle{\bf r}_{pn}^2\rangle=\int d^3{\bf r}_{\Lambda}\int d^3{\bf r}_{pn}{\bf r}^2_{pn}\left|\phi_{^3_\Lambda \rm H}({\bf r}_{pn},{\bf r}_\Lambda)\right|^2=3b_{pn}^2$. Similar goes for the pn-$\Lambda$ distance $\langle {\bf r}_{\Lambda}^2\rangle=\int d^3{\bf r}_{\Lambda}\int d^3{\bf r}_{pn}{\bf r}^2_{\Lambda}\left|\phi_{^3_\Lambda \rm H}({\bf r}_{pn},{\bf r}_\Lambda)\right|^2=\frac{9}{4}b_\Lambda^2$.} $\sqrt{\langle {\bf r}_{pn}^2\rangle}=\sqrt{3}b_{pn}\approx3$~fm and $\sqrt{\langle {\bf r}_{\Lambda}^2\rangle}=\frac{3}{2}b_\Lambda\approx10.8^{+3}_{-1.5}$~fm. It is important to note, however, that Ref.~\cite{Hildenbrand:2019sgp} reported the full numerical momentum space form factors and the $\sqrt{\langle{\bf r}^2\rangle}$ parameters they quote refer only to the small-${\bf k}^2$ expansion of these form factors. A more accurate treatment of the wave function is obtained by using the full form factors, which we do in Sec.~\ref{ss:waveacc}. Based on that analysis, for the Gaussian approximation we set $b_{pn}=1.52$~fm (as opposed to $b_{pn}=1.73$~fm that would be read from the low-${\bf k}^2$ fit) and $b_\Lambda=7.2^{+2}_{-1}$~fm.

With the Gaussian nuclear wave functions described in this section, assuming the HBT source parameterisation of Eqs.~(\ref{eq:C2ssng}-\ref{eq:C3ssng}) and recalling the spins $s_N=s_{\rm ^3He}=s_{\rm ^3_\Lambda H}=\frac{1}{2},\;s_d=1$, the coalescence factors evaluated from Eqs.~(\ref{eq:B2mom}),~(\ref{eq:B3Hmom}) and~(\ref{eq:B3mom}) are given by\footnote{Our Eq.~(\ref{eq:B2G}) is consistent with Eq.~(28) of~\cite{Blum:2019suo}. Our Eq.~(\ref{eq:B3G}) corrects a typo in Eq.~(31) of~\cite{Blum:2019suo}, where one should replace $(d_A/2)\to(d_A/\sqrt{2})$.}:
\be\label{eq:B2G}\B_2&\approx&\frac{12\pi^{\frac{3}{2}}\lambda_2}{m\left(b_d^2+4R_l^2\right)^{\frac{1}{2}}\left(b_d^2+4R_o^2\right)^{\frac{1}{2}}\left(b_d^2+4R_s^2\right)^{\frac{1}{2}}},\\
\label{eq:B3G}\B_3&\approx&\frac{16\pi^3\lambda_3}{\sqrt{3}m^2\left(b_{\rm ^3He}^2+2R_l^2\right)\left(b_{\rm ^3He}^2+2R_o^2\right)\left(b_{\rm ^3He}^2+2R_s^2\right)},\\
\label{eq:B3HG}\B_{3\Lambda}&\approx&\frac{16\pi^3\lambda_{3\Lambda}}{\sqrt{3}m^2\left(b_{pn}^2+2R_l^2\right)^{\frac{1}{2}}\left(b_{\Lambda}^2+2R_l^2\right)^{\frac{1}{2}}\left(b_{pn}^2+2R_o^2\right)^{\frac{1}{2}}\left(b_{\Lambda}^2+2R_o^2\right)^{\frac{1}{2}}\left(b_{pn}^2+2R_s^2\right)^{\frac{1}{2}}\left(b_{\Lambda}^2+2R_s^2\right)^{\frac{1}{2}}}.
\ee

We stress again that the analytic results obtained with Gaussian wave functions are brought here as means for an easy, rough assessment of the coalescence factor. More accurate calculations should use more accurate wave functions, especially for the D and ${\rm ^3_\Lambda H}$. We consider this refinement next.

\subsection{Nucleus wave functions: more accurate parameterisation.}\label{ss:waveacc}
Here we consider more accurate parameterisations for the wave functions of the D and the ${\rm ^3_\Lambda H}$. For $^3$He we maintain the Gaussian ansatz of Sec.~\ref{ss:waveGaus}.

\subsubsection{Deuteron wave function.}
A more accurate parametrisation of the D wave function, that should be used instead of the Gaussian ansatz for quantitative analyses, is given by the Hulthen formula:
\be\label{eq:Dhul}\phi_d({\bf r})&=&\sqrt{\frac{\alpha\beta(\alpha+\beta)}{2\pi(\alpha-\beta)^2}}\frac{e^{-\alpha|{\bf r}|}-e^{-\beta|{\bf r}|}}{|{\bf r}|}.\ee
The RMS radius is given by 
\be r_{rms}^2&=&\frac{\beta(\alpha +\beta )}{8\alpha^2 (\alpha -\beta )^2} \left(1-\frac{16\alpha^3}{(\alpha +\beta )^3}+\frac{\alpha^3}{\beta ^3}\right)=\frac{1}{8\alpha^2}\left(1+\frac{3\alpha}{\beta}+\mathcal{O}\left(\frac{\alpha^2}{\beta^2}\right)\right).\ee
For this parameterisation it is assumed that $\beta>\alpha$, so above we expanded in the ratio $(\alpha/\beta)$. For the numerical evaluation we set $\alpha=0.2$~fm$^{-1}$ and $\beta=1.56$~fm$^{-1}$, reproducing $r_{rms}=2.13$~fm~\cite{Sick:2015spa}. We have checked that using the slightly different values of $\alpha$ and $\beta$ quoted in Ref.~\cite{Lamia:2012zz} gives results that are equal to ours to 5\% accuracy. 
The form factor Eq.~(\ref{eq:D(k)}) can only be obtained numerically. Amusingly, in the 1D Gaussian source limit Eq.~(\ref{eq:C2ss}) we do not need it because the coalescence factor itself can be obtained analytically:
\be\label{eq:B2hul}\B_2&=&\frac{3}{2m}\int d^3{\bf k}\,e^{-{\bf k}^2R_{\rm inv}^2}\int d^3{\bf r}\,|\phi_d({\bf r})|^2\,e^{-i{\bf kr}}\\
&=&\frac{3\pi^2}{mR_{\rm inv}^2}\frac{\alpha  \beta  (\alpha +\beta )}{(\alpha -\beta )^2} \left(e^{4 \alpha ^2 R_{\rm inv}^2} \text{erfc}\left(2 \alpha  R_{\rm inv}\right)-2 e^{(\alpha +\beta )^2 R_{\rm inv}^2} \text{erfc}\left((\alpha +\beta ) R_{\rm inv}\right)+e^{4 \beta ^2 R_{\rm inv}^2} \text{erfc}\left(2 \beta  R_{\rm inv}\right)\right).\no\ee

\begin{figure}[htbp]
  \begin{center}
   \includegraphics[width=0.6\textwidth]{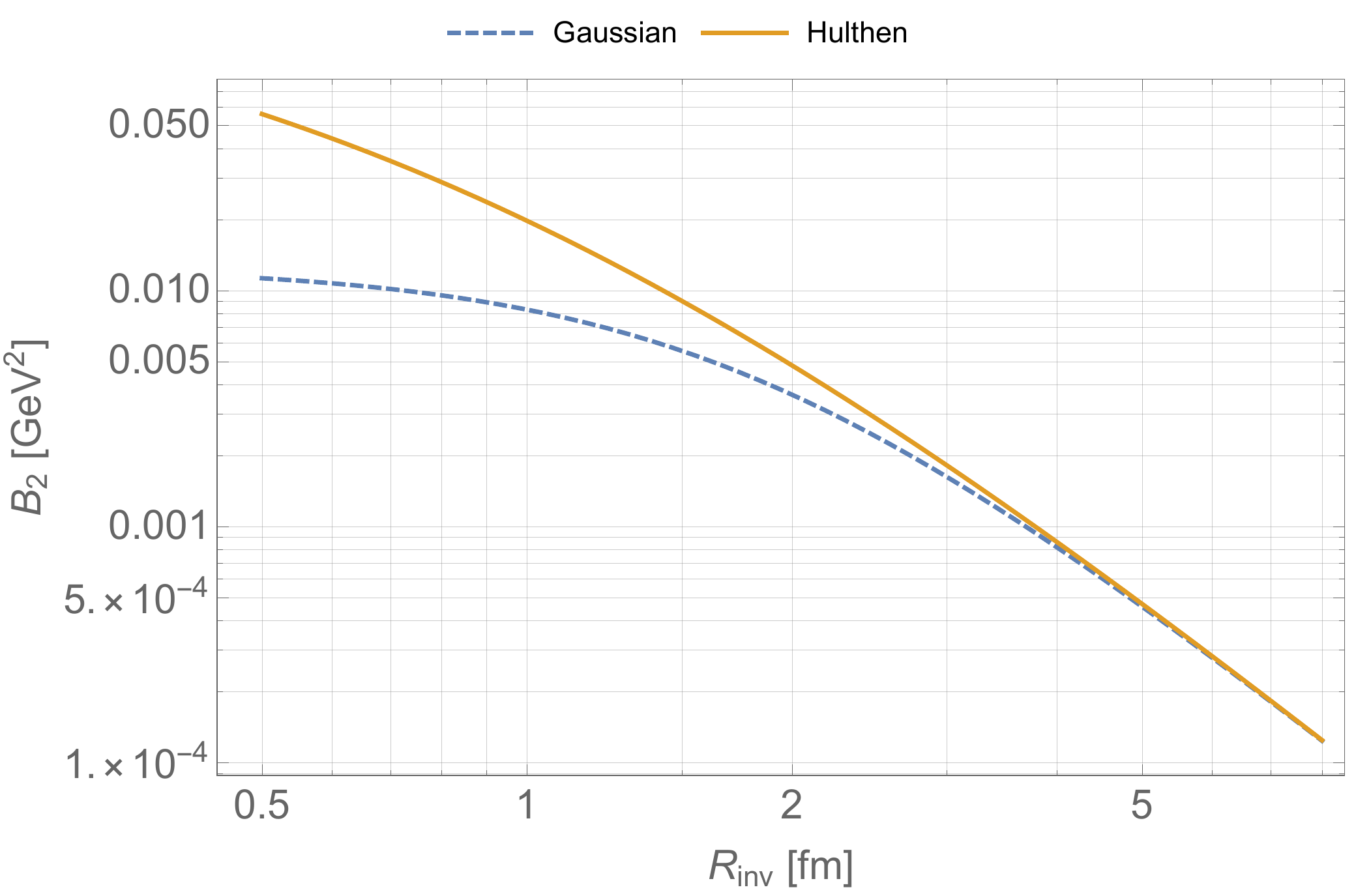}
  \end{center}
  \caption{Deuteron coalescence factor $\B_2$, calculated in the isotropic (1D) Gaussian source model, showing the difference between the Gaussian and Hulthen wave function parameterisations.}
  \label{fig:B2}
\end{figure}

The coalescence factor $\B_2$ is shown in Fig.~\ref{fig:B2}. Solid line shows the prediction of Eq.~(\ref{eq:B2hul}), obtained for the Hulthen wave function. For comparison, dashed line shows the less accurate Gaussian wave function prediction of Eq.~(\ref{eq:B2G}). For simplicity here we use the isotropic 1D Gaussian source model with $R_o=R_s=R_l=R_{\rm inv}$ and $\lambda_2=1$. Note that measurements at small $R_{\rm inv}\sim1$ are possible in pp or pPb collisions, and may be sensitive to effect of the wave function seen in Fig.~\ref{fig:B2}.

To use the 3D source Eq.~(\ref{eq:C2ssng}), the coalescence factor needs to be calculated numerically from Eq.~(\ref{eq:B2mom}) or~(\ref{eq:B2}).

\subsubsection{Hypertriton wave function.}\label{sss:HTwave}
\begin{figure}[htbp]
  \begin{center}
   \includegraphics[width=0.495\textwidth]{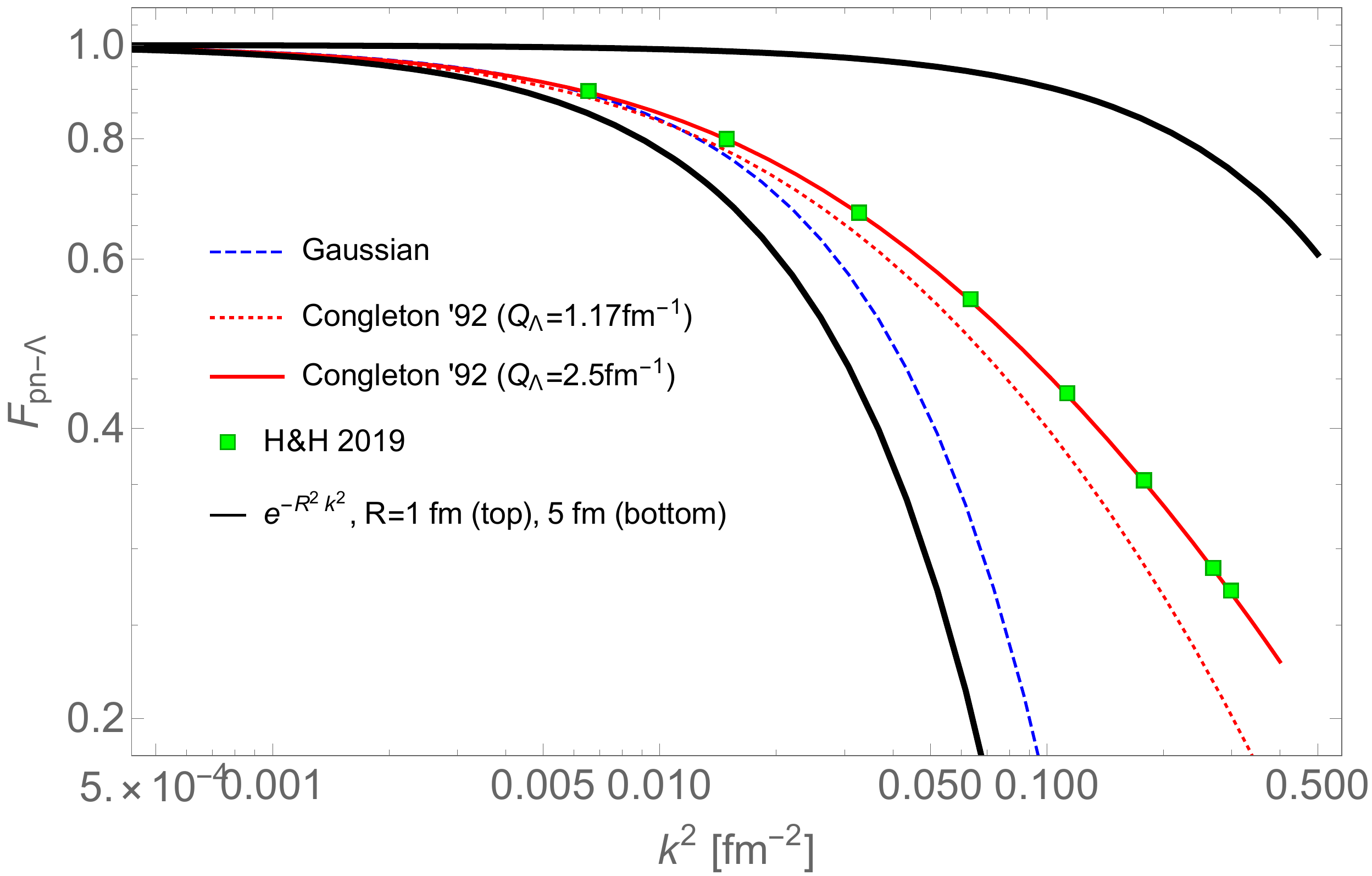}
   \includegraphics[width=0.495\textwidth]{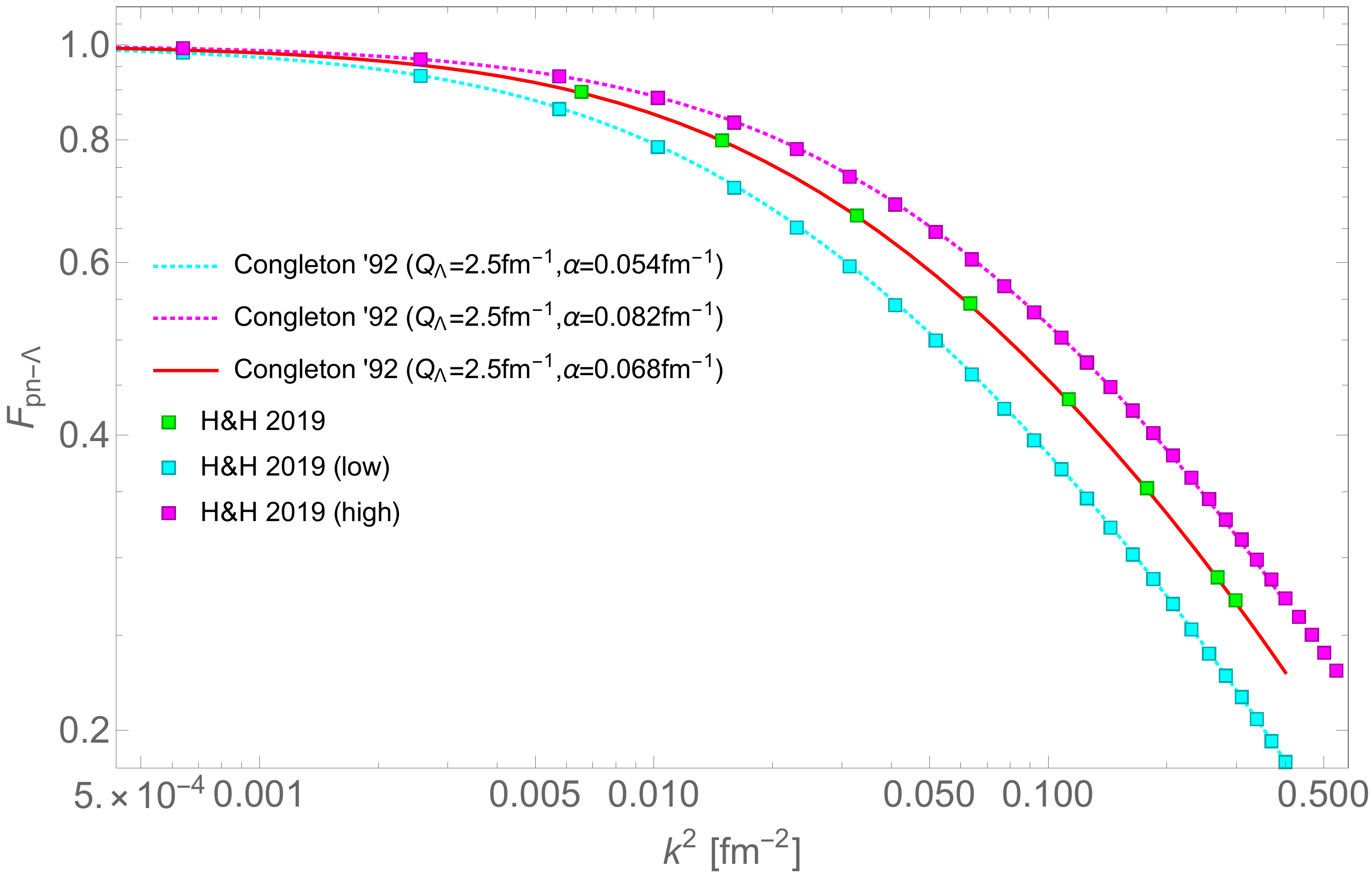}
  \end{center}
  \caption{Hypertriton $pn-\Lambda$ form factor. Green squares show the result from the three-body calculation of Hildenbrand \& Hammer~\cite{Hildenbrand:2019sgp}. {\bf Left:} Dotted and solid red line shows the effective two-body calculation of Congleton~\cite{Congleton:1992kk} for different values of their $Q_\Lambda$ parameter. Blue dashed line shows a Gaussian approximation with the same charge radius as found in~\cite{Hildenbrand:2019sgp}. For comparison, thick solid black lines show exponential factors with scale radius $R=1$~fm (top) and $5$~fm (bottom), respectively. {\bf Right:} Cyan and magenta markers show a numerical calculation delimiting the uncertainty due to the $^3_\Lambda$H binding energy, as implemented in~\cite{Hildenbrand:2019sgp}.}
  \label{fig:DLamk2}
\end{figure}

Hildenbrand \& Hammer~\cite{Hildenbrand:2019sgp} reported a three-body theoretical calculation of the ${\rm ^3_\Lambda H}$ wave function. 
Green squares in the {\bf left panel} of Fig.~\ref{fig:DLamk2} show the projected $pn-\Lambda$ form factor obtained in that work. It is well reproduced by the two-body calculation of Congleton~\cite{Congleton:1992kk}, shown by solid red line, provided we adjust the $Q_\Lambda$ parameter of~\cite{Congleton:1992kk} from $Q_\Lambda=1.17$~fm$^{-1}$ in the original paper to $Q_\Lambda=2.5$~fm$^{-1}$. For simplicity we therefore consider the effective $pn-\Lambda$ wave function from~\cite{Congleton:1992kk}, given in momentum space by
\be\hat\phi_{{\rm ^3_\Lambda H}(\Lambda d)}({\bf q})&=&A\frac{e^{-\frac{{\bf q}^2}{Q_\Lambda^2}}}{{\bf q}^2+\alpha_\Lambda^2}\;=\;\int d^3{\bf r}_\Lambda e^{-i{\bf r}_\Lambda{\bf q}_{\Lambda}}\phi_{{\rm ^3_\Lambda H}(\Lambda d)}({\bf r}_\Lambda)\ee
where the normalisation constant $A$ is defined such that $\int d^3{\bf r}_\Lambda \left|\phi_{{\rm ^3_\Lambda H}(\Lambda d)}({\bf r}_\Lambda)\right|^2=\int \frac{d^3{\bf q}}{(2\pi)^3}\left|\hat\phi_{{\rm ^3_\Lambda H}(\Lambda d)}({\bf q})\right|^2=1$. 
Using the $pn-\Lambda$ wave function of~\cite{Congleton:1992kk}, with $Q_\Lambda=2.5$~fm$^{-1}$ and $\alpha=0.068$~fm$^{-1}$, we can calculate the integrals in Eq.~(\ref{eq:B3Hmom}) or~(\ref{eq:B3H}) numerically. 

In the {\bf right panel} of Fig.~\ref{fig:DLamk2} we show numerical calculations from~\cite{Hildenbrand:2019sgp}, delimiting the uncertainty due to the $^3_\Lambda$H binding energy\footnote{We are grateful to Fabian Hildenbrand for providing this calculation to us.}. We find that the lower (upper) range for the form factor is well fitted again by Congleton's formula, with $Q_\Lambda=2.5$~fm$^{-1}$ and $\alpha=0.054$ (0.082)~fm$^{-1}$, respectively.

\begin{figure}[htbp]
  \begin{center}
   \includegraphics[width=0.55\textwidth]{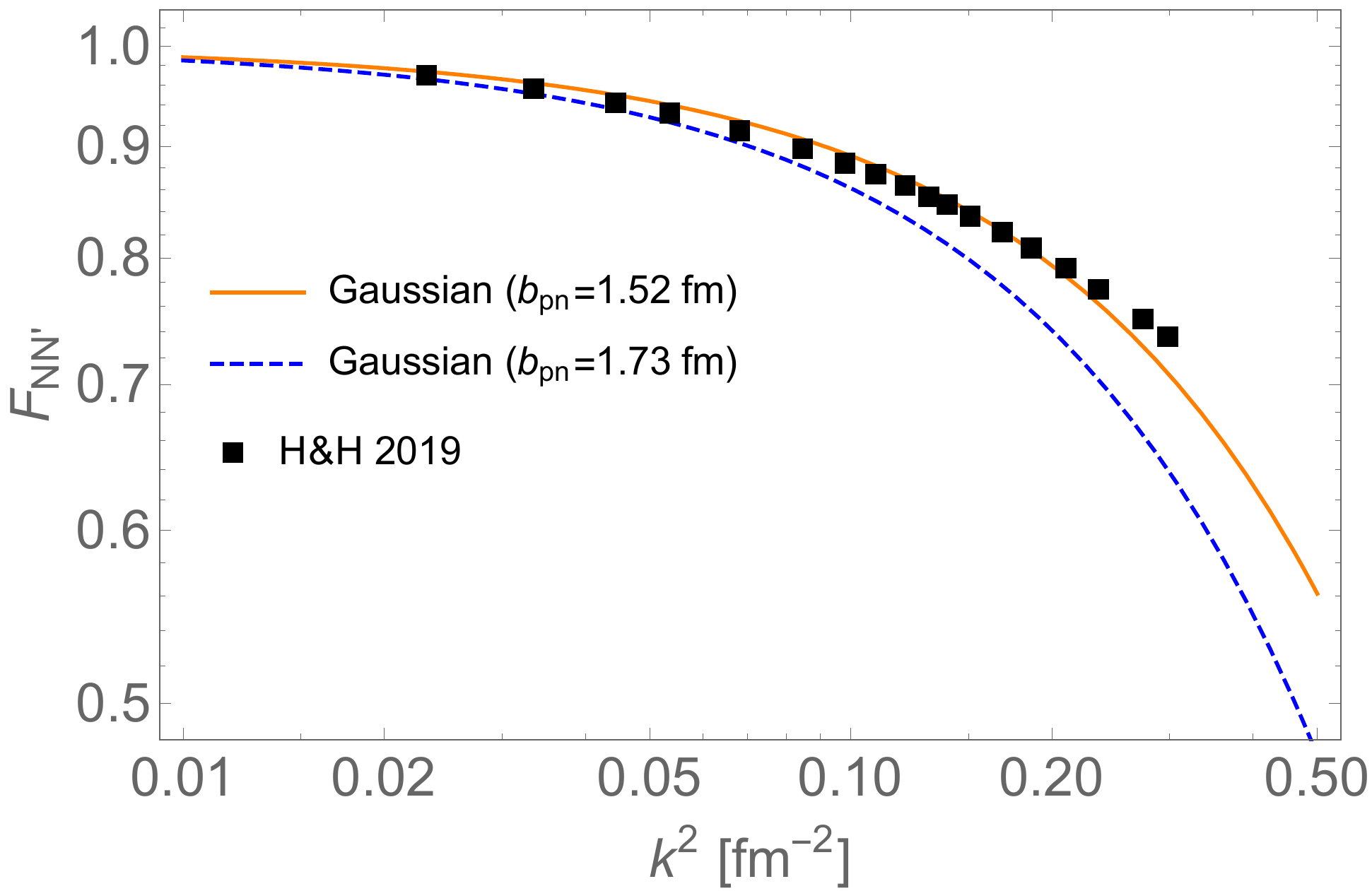}
  \end{center}
  \caption{Hypertriton $pn$ form factor. Black squares show the result from the three-body calculation of Hildenbrand \& Hammer~\cite{Hildenbrand:2019sgp}. Solid orange and dashed blue lines show a Gaussian approximation a-la Eq.~(\ref{eq:FHTGauss}) with $b_{pn}=1.52$~fm and $b_{pn}=1.73$~fm, respectively.}
  \label{fig:FNN}
\end{figure}

Ref.~\cite{Hildenbrand:2019sgp} also provided the effective form factor for the pn subsystem. Their numerical result is shown by black squares in Fig.~\ref{fig:FNN}. This form factor is reasonably well reproduced by a Gaussian of the form given by Eq.~(\ref{eq:FHTGauss}), with $b_{pn}=1.52$~fm (orange solid line). This can be compared with the low-${\bf k}^2$ expansion of the form factor, which would lead to $b_{pn}({\rm low-{\bf k}^2~fit})=1.73$~fm as noted in Sec.~\ref{ss:waveGaus}. Given this discussion we can maintain the Gaussian ansatz of the pn factor in Eq.~(\ref{eq:FHTGauss}), setting $b_{pn}=1.52$~fm. 

\begin{figure}[htbp]
  \begin{center}
   \includegraphics[width=0.6\textwidth]{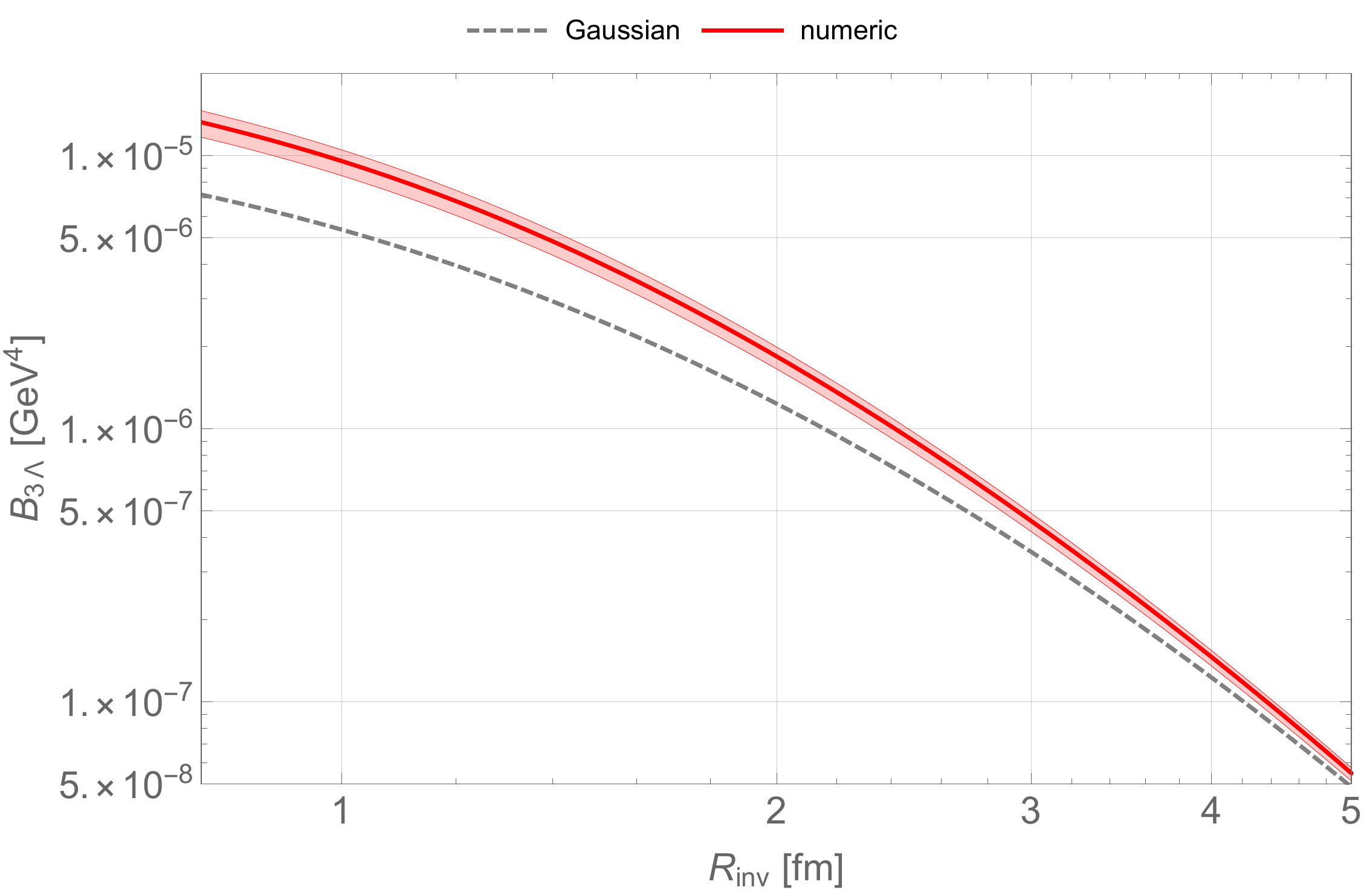}
  \end{center}
  \caption{Hypertriton coalescence factor $\B_{3\Lambda}$, calculated in the isotropic (1D) Gaussian source model, showing the difference between the Gaussian (dashed grey) and the more realistic numerical (red) wave function parameterisations.}
  \label{fig:B3LGaussNum}
\end{figure}

In Fig.~\ref{fig:B3LGaussNum} we plot $\B_{3\Lambda}$ vs $R_{\rm inv}$, showing the difference between the Gaussian (dashed grey) and the more realistic numerical (red) wave function parameterisations. Shaded band around the numerical result reflects the wave function uncertainty, as depicted in the right panel of Fig.~\ref{fig:DLamk2}.

\section{Comparison with data.}\label{s:dat}

\subsection{Deuteron and $^3$He.}\label{ss:DHe3dat}
\begin{figure}[htbp]
  \begin{center}
   \includegraphics[width=0.495\textwidth]{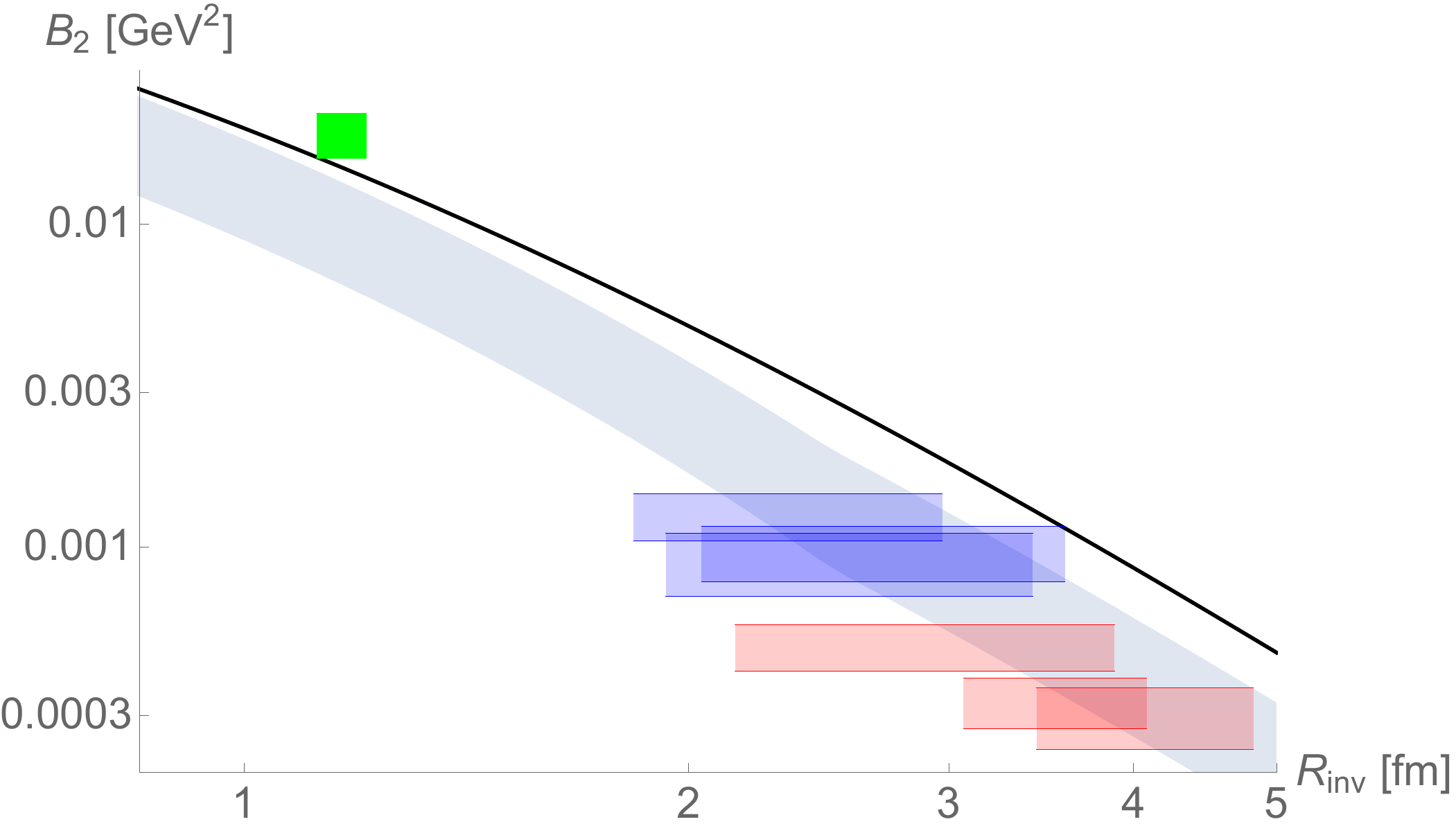}
    \includegraphics[width=0.495\textwidth]{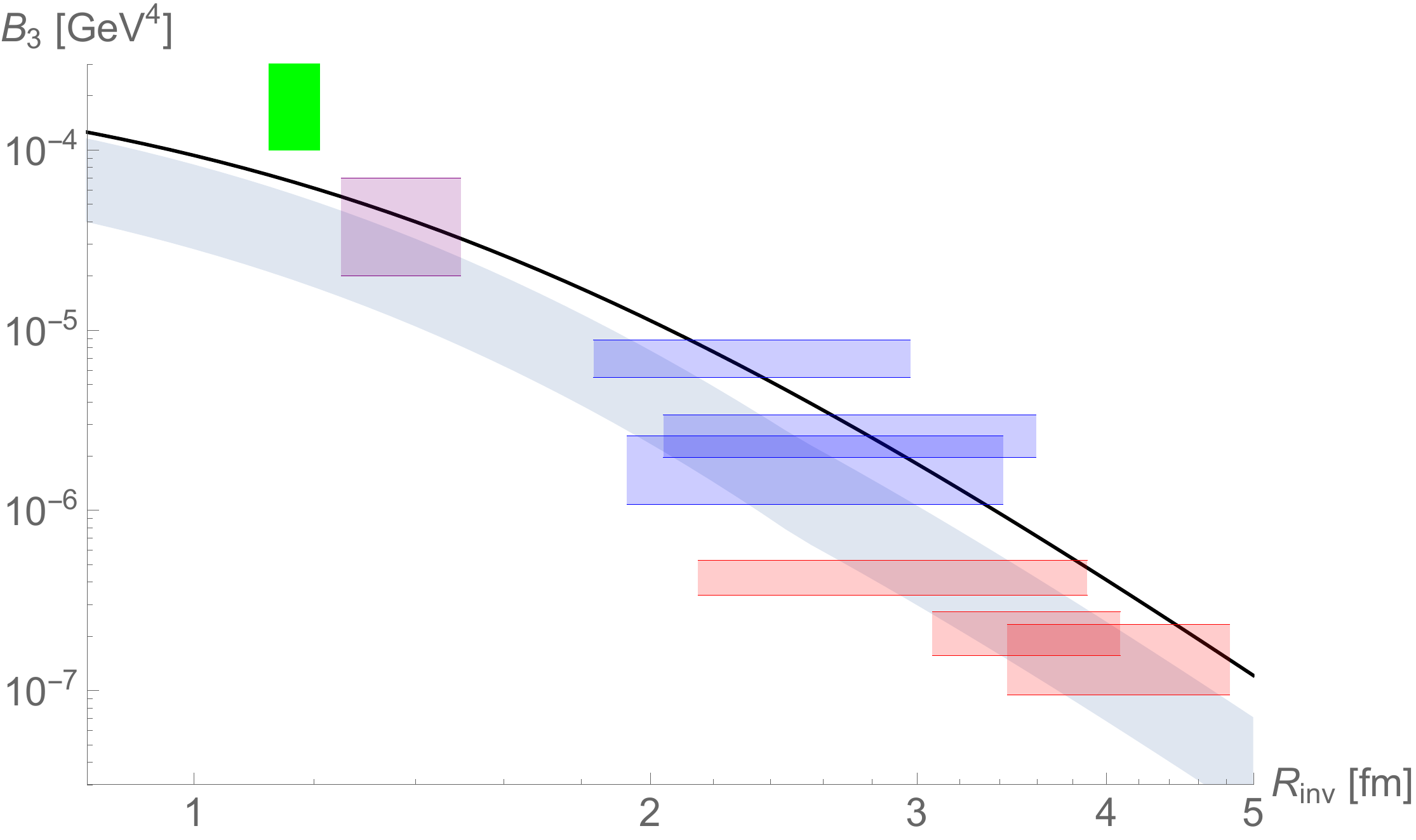}
    \end{center}
  \caption{Summary of D and $^3$He data, reproduced from Ref.~\cite{Blum:2019suo} with some improvements (see text). {\bf Left:} $\mathcal{B}_2$ vs. $R_{\rm inv}$. {\bf Right:} $\mathcal{B}_3$ vs. $R_{\rm inv}$.}
  \label{fig:BR}
\end{figure}

Fig.~\ref{fig:BR} shows the theoretical prediction for $\mathcal{B}_2$ ({\bf left}) and $\mathcal{B}_3$ ({\bf right}), calculated as function of the 1D HBT parameter $R_{\rm inv}$ using  Eq.~(\ref{eq:B2hul}) (based on the Hulthen wave function) for D and Eq.~(\ref{eq:B3G}) (Gaussian wave function) for $^3$He, with $R_o=R_s=R_l=R_{\rm inv}$. 
The calculation, shown by a grey shaded band, uses an estimate of the experimentally measured value of $\lambda_2$. To define the upper edge of the bands, we interpolate between $\lambda_2=\{1,0.7,0.7\}$ defined at $R_{\rm inv}=\{0.85,2.5,5\}$. To define the lower edge we interpolate between $\lambda_2=\{0.5,0.3,0.3\}$ defined at $R_{\rm inv}=\{0.85,2.5,5\}$. This range of $\lambda_2$ is estimated as follows. For large $R_{\rm inv}$ values we have a measurement of $\lambda_2$ in PbPb collisions~\cite{Adam:2015vja}, obtained as the sum $\lambda_2\approx\lambda^{(pp)}+\lambda^{(p\Lambda)}$ in the notation of that paper. For small $R_{\rm inv}$ values, corresponding to pp collisions, we have no measurement of $\lambda_2$; Refs.~\cite{Acharya:2018gyz,Abelev:2012sq}, which could in principle measure $\lambda_2$, effectively fixed $\lambda_2\to1$ in their fit. As a next best solution we use $\lambda_2$ measured from kaon femtoscopy~\cite{Abelev:2012sq}. This is a potentially reasonable estimate because Ref.~\cite{Adam:2015vja} demonstrated HBT parameters that were the same, within measurement uncertainties, for kaon and proton final states at the same $m_t$. 
In addition to this attempt to estimate $\lambda_2$ from data we also show the result obtained fixing $\lambda_2=1$ (black solid line).

Based on our analysis in App.~\ref{sss:model} we can expect that for the $p_t\lesssim1.5$~GeV values, in which the cluster data in Fig.~\ref{fig:BR} are given, our use of the simplistic 1D source parameterisation should cause us to over-estimate a more accurate 3D prediction (not available to us, as the experiments reported 1D HBT fits only) by $\lesssim20\%$ or so. We do not include this uncertainty in the plot. It adds up other sources of systematic uncertainty, expected to be roughly at a similar level (with unknown signs), due to the smoothness, equal-time, and factorisation approximations (the latter relevant for $^3$He only).

The comparison to experimental data is as follows. The red horizontal bands in Fig.~\ref{fig:BR} show experimental coalescence factor measurements for PbPb at (0-10\%) (for $\mathcal{B}_2$) and (0-20\%) (for $\mathcal{B}_3$) centrality classes~\cite{Adam:2015vda}. Each of the three red bands corresponds to a different bin in $m_t$, among the three bins shown in the HBT $R_{\rm inv}$ measurement~\cite{Adam:2015vja}. The blue horizontal bands show the result for the (20-40\%) (for $\mathcal{B}_2$) and (20-80\%) (for $\mathcal{B}_3$) events, respectively, again from~\cite{Adam:2015vda}. 

The green band shows the result for p-p collisions~\cite{Acharya:2017fvb}\footnote{We thank Bhawani Singh and the Fabbietti TUM group for pointing out a typo in the plot of the $\B_2$ data for pp collisions in Ref.~\cite{Blum:2019suo}.}.

For $^3$He we can also add a crudely estimated data point for pPb collisions. To do so, we combine the 1D HBT $R_{\rm inv}$ measurement of kaon femtoscopy reported in~\cite{Acharya:2019fip} with the $^3$He measurement of~\cite{Acharya:2019xmu}. To approximately match $m_t$ between the data sets, we use here the highest $k_t$ bin in~\cite{Acharya:2019fip} and the lowest $p_t$ bin in~\cite{Acharya:2019xmu}. We use the (0-20\%) multiplicity class from~\cite{Acharya:2019fip}, joining together the (0-10\%) and (10-20\%) $\B_3$ data from~\cite{Acharya:2019xmu}. The result is shown in purple in the right panel of Fig.~\ref{fig:BR}.

As should be clear by now, a main uncertainty in the theory prediction shown in Fig.~\ref{fig:BR} is related to the determination of the chaoticity parameters $\lambda_2$ and $\lambda_3$. A measurement of $\lambda_2$ is missing for the pp and pPb systems, and we had to complete this information by using results from kaon femtoscopy. While the HBT results for kaons and protons come close to each other where they are available at the same value of $m_t$, it is clear that a proton HBT result is more suitable for the derivation of nuclear coalescence. For $\lambda_3$ we have no data, as such data would require three-proton femtoscopy. In Fig.~\ref{fig:BR} we bypassed this by assuming $\lambda_3=\lambda_2^{\frac{3}{2}}$. Addressing this issue experimentally would be challenging, and we do not know of a model-independent way to estimate the associated theory uncertainty. Assessing the uncertainty within specific HXS models, along the lines of App.~\ref{sss:model}, may be warranted in future work.

Another obvious difficulty is due to the need to construct Fig.~\ref{fig:BR} patch-wise from data at different, often only partially overlapping, multiplicity class and $p_t$ or $m_t$ bins. A dedicated experimental analysis combining HBT and cluster yields would solve this problem.

Altogether, Fig.~\ref{fig:BR} shows that coalescence is roughly consistent with the D and $^3$He data for systems ranging from pp to pPb and PbPb at different regions of $p_t$ and at different multiplicity classes. This comparison spans a dynamical range of about a factor of 30 for $\B_2$ and a factor of $10^3$ for $\B_3$. While, as we discussed, there are theoretical and experimental uncertainties, there are no free parameters once HBT calibrates the computation. From this point of view, the usual claim to fame of the SHM~\cite{Andronic:2017pug}, to describe the yields of nuclei across many orders of magnitude, is seen to be comparably well applicable to coalescence.

Having said that, it is worth highlighting that for pp collisions the experimental coalescence factors for both D and $^3$He are found to be higher than the coalescence prediction. Depending mainly on the poorly determined value of $\lambda_2$, but also on possible systematic uncertainties related to different event classes entering the HBT and cluster measurements, the discrepancy could be as much as a factor of 2 for D and a factor of 4 for $^3$He. We think that this situation is strong motivation for a joint experimental analysis of coalescence and HBT in small systems.

\subsection{Hypertriton.}\label{ss:HTdat}
\begin{figure}[htbp]
  \begin{center}
   \includegraphics[width=0.475\textwidth]{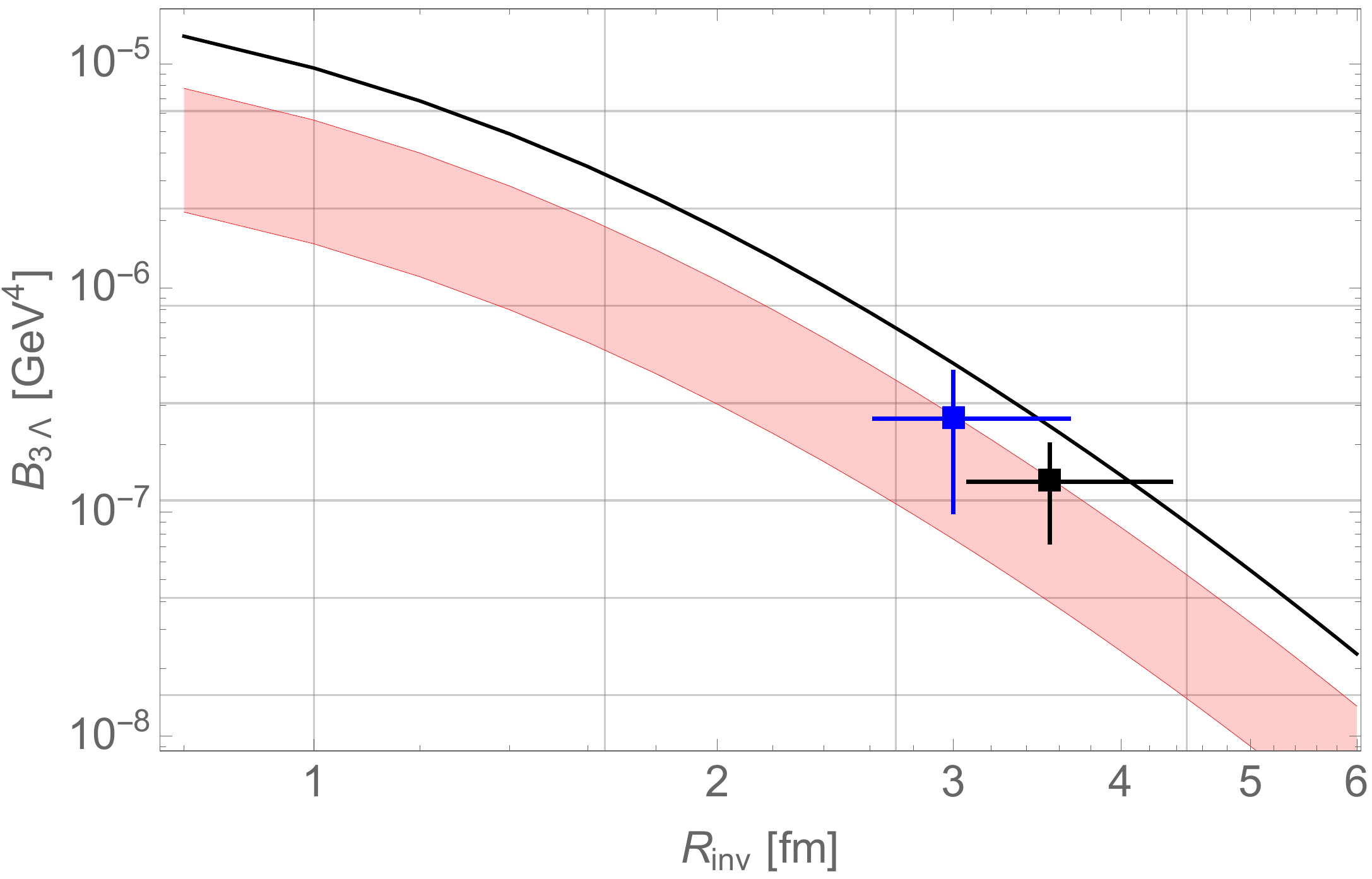}
   \includegraphics[width=0.5\textwidth]{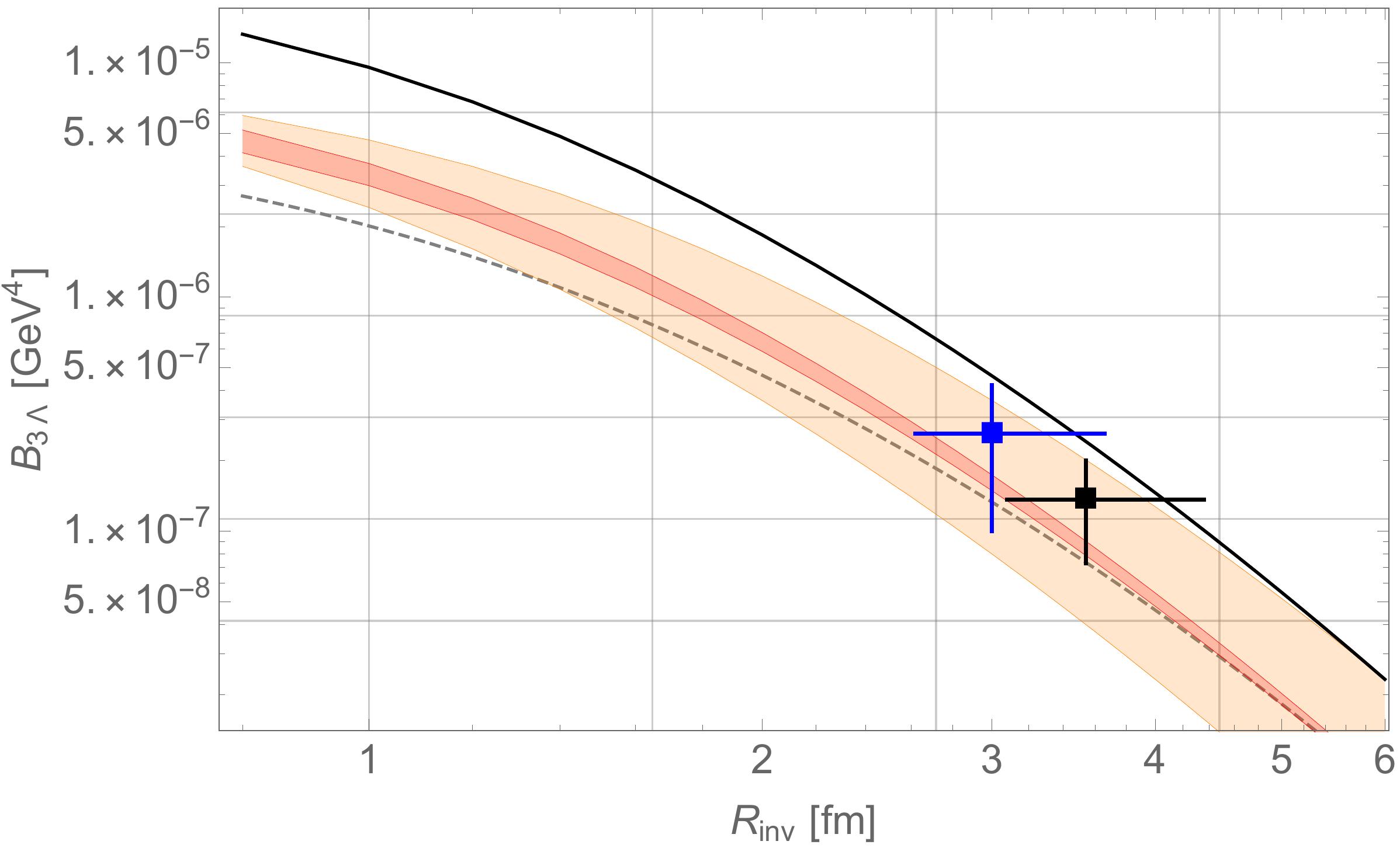}
  \end{center}
  \caption{$\B_{3\Lambda}$ vs. the 1D HBT radius $R_{\rm inv}$. {\bf Left:} The coalescence prediction for the numerical wave function of Sec.~\ref{sss:HTwave}, using $\lambda_3=\lambda_2^{\frac{3}{2}}$ taken from data (red band). The width of the band derives from the uncertainty on $\lambda_2$. For reference, the result for $\lambda_2=1$ is shown by solid black line. {\bf Right:} Impact of wave function and $R_{\rm inv}$ uncertainties. Red band: numerical wave function uncertainty. Orange band: $R_{\rm inv}$ uncertainty (see text). Grey dashed: Gaussian wave function approximation. Black line is same as on the left.
  }
  \label{fig:B3L}
\end{figure}

Ref.~\cite{Adam:2015yta} reported a measurement of $^3_\Lambda$H in PbPb collisions. After some gymnastics we extract the measured $\B_{3\Lambda}$ from the three $p_t$ bins in the left panel of their Fig.~7. For the bins\footnote{Our $p_t$ is corresponds to $p_t/A$ as defined in~\cite{Adam:2015yta}.} $p_t\approx(0.66-1.34),\,(1.34-2),$ and $(2-2.34)$~GeV we find $\B_{3\Lambda}\approx(7.3-20)\times10^{-8},\,(1-4.2)\times10^{-7}$, and $(1.7-12)\times10^{-5}$~GeV$^4$, respectively. The multiplicity class is (0-10\%). For HBT data, we have Ref.~\cite{Adam:2015vja} with $R_{\rm inv}$ and $\lambda_2$ measured in the same multiplicity class, but binned in $m_t$ rather than $p_t$ (see Figs.~7-8 in~\cite{Adam:2015vja}). We can match the first low $p_t$ bin of~\cite{Adam:2015yta} into the $m_t$ range covered in~\cite{Adam:2015vja}. 
We also consider the second $p_t$ bin of~\cite{Adam:2015yta} that somewhat overshoots the coverage of the last $m_t$ bin in~\cite{Adam:2015vja}. With some interpolation (and, for the second bin, extrapolation) of the results from~\cite{Adam:2015vja}, we obtain the corresponding estimated ranges in $R_{\rm inv}$.   
The result is shown by markers in Fig.~\ref{fig:B3L}. 

To plot the theory prediction, we need an estimate of $\lambda_{3\Lambda}$. On the {\bf left} panel of Fig.~\ref{fig:B3L} we set $\lambda_{3\Lambda}=\lambda_2^{\frac{3}{2}}$, with $\lambda_2$ taken in the range $(0.3-0.7)$, estimated from the data~\cite{Adam:2015vja}. The theory prediction is shown by the red band, with the width of the band reflecting the allowed range in $\lambda_2$. In this exercise we use the numerical wave function of Sec.~\ref{sss:HTwave}. For reference, the result for $\lambda_2=1$ is shown by a black line. 
On the {\bf right} panel we illustrate the wave function uncertainty, as well as the uncertainty (mentioned earlier in Sec.~\ref{ss:GS}) associated with determining $R_{\rm inv}$ from pp vs. p$\Lambda$ or $\Lambda\Lambda$ correlations. Setting $\lambda_2=0.5$ (in between 0.3 and 0.7), the result of varying the numerical wave function in the range corresponding to the right panel of Fig.~\ref{fig:DLamk2} is shown by the red band. To implement the $R_{\rm inv}$ uncertainty, an orange band shows the effect of shifting the $R_{\rm inv}$ argument, entering the $\B_{3\Lambda}$ computation, by $\pm20\%$ w.r.t. to the $R_{\rm inv}$ value on the x-axis. For comparison we also show the Gaussian wave function Eq.~(\ref{eq:B3HG}) in grey dashed. Black line for reference is same as on the left. 

\begin{figure}[htbp]
  \begin{center}
   \includegraphics[width=0.6\textwidth]{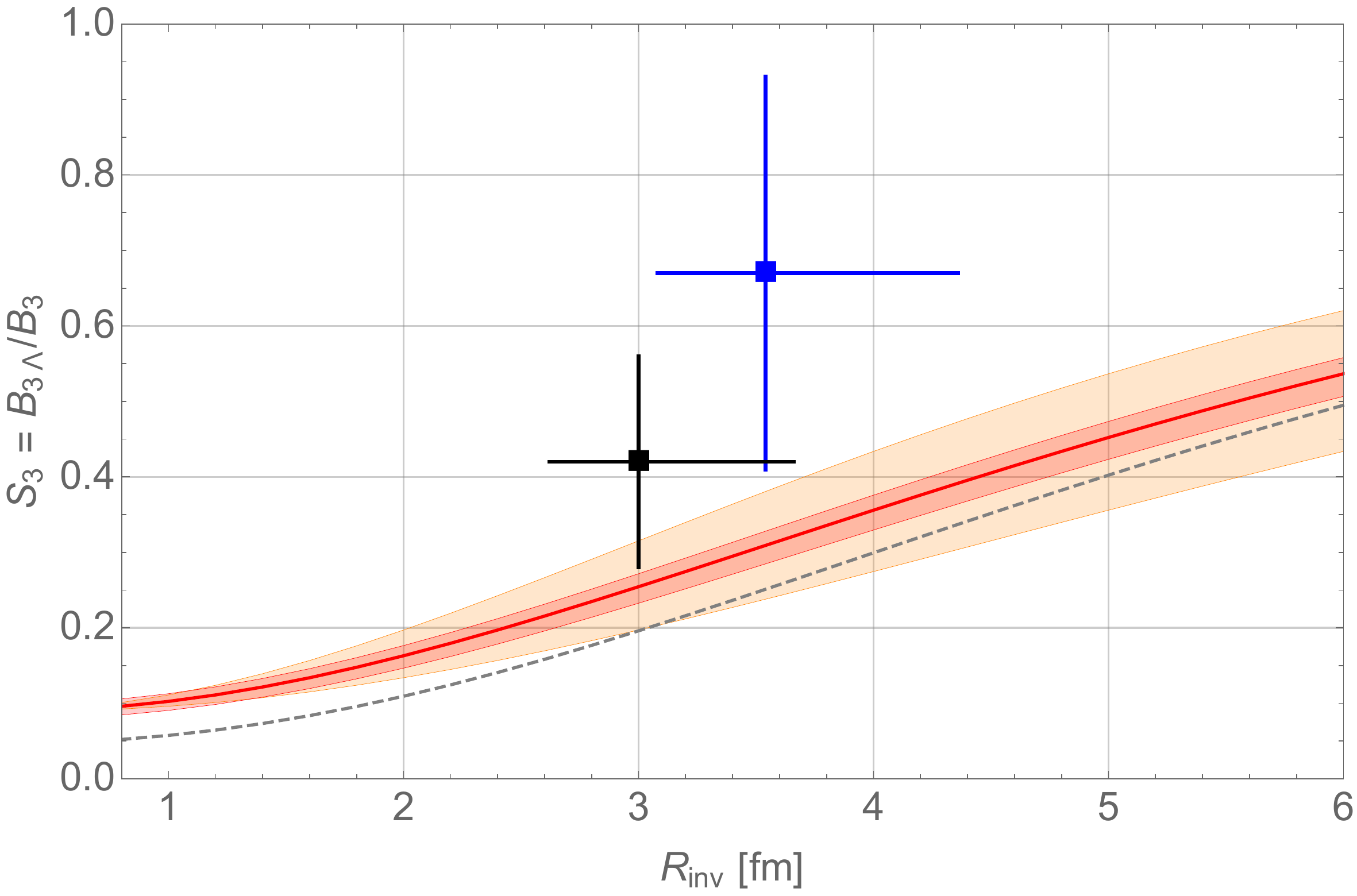}
  \end{center}
  \caption{$S_3$ vs. $R_{\rm inv}$ predicted in the coalescence model compared with data.
  }
  \label{fig:S3}
\end{figure}

Comparing between the left and right panels of Fig.~\ref{fig:B3L} we can see that the estimated uncertainty due to the chaoticity parameter $\lambda_3$ (or $\lambda_2$) is larger than that due to the $^3_\Lambda$H wave function, again motivating an experimental effort to extract the chaoticity from data. We can also try to bypass some of this uncertainty as follows. If Nature is kind, and $\lambda_3\approx\lambda_{3\Lambda}$, then the uncertainty associated with $\lambda_{3\Lambda}$ may cancel in the ratio\footnote{Not to be confused with the three-particle normalised source, $\mathcal{S}_3$, defined in Sec.~\ref{sss:HTHe3}.}:
\be S_3&=&\frac{\B_{3\Lambda}}{\B_3}.\ee
For the first $p_t$ bin of~\cite{Adam:2015yta} we find $\B_3\approx(1.4-2.7)\times10^{-7}$~GeV$^4$, and for the second bin $\B_3\approx(3.3-6.9)\times10^{-7}$~GeV$^4$. We could, in principle, combine this directly with $\B_{3\Lambda}$ to extract the measured $S_3$. However, this would ignore the fact that some of the experimental uncertainty involved in deriving $\B_{3}$ and $\B_{3\Lambda}$ could cancel out in the ratio. 
Instead, we therefore adopt the following procedure. Taking the hypertriton spectrum from~\cite{Adam:2015yta} (averaging the $^3_\Lambda$H and $\overline{^3_\Lambda{\rm H}}$ results), we divide by the $^{3}$He-spectrum from~\cite{Adam:2015vda} and scale by the $\Lambda$/p ratio taken from~\cite{Abelev:2013xaa,Abelev:2013vea}, evaluated in the corresponding $p_t$ interval. The result is shown by markers in Fig.~\ref{fig:S3}.

The theoretical prediction for $S_3$, using the numerical $^3_\Lambda$H wave function of Sec.~\ref{sss:HTwave}, is shown by the red band in Fig.~\ref{fig:S3} with the band width determined by the wave function uncertainty. The effect of varying $R_{\rm inv}$ by $\pm20\%$ is shown by the orange band. 
For comparison we also show the Gaussian approximation with a grey dashed line.
The coalescence prediction is somewhat below the data, with tension at the $\sim2\sigma$ level\footnote{To quantify the significance of the tension more precisely we would need to combine the two data points, which very much overlap in $R_{\rm inv}$. We prefer to leave these details to a dedicated experimental analysis.}. It is clear from Fig.~\ref{fig:S3} that a more precise experimental measurement of $S_3$ in conjunction with HBT is a promising observable to exclude (or support) the framework.  
At small $R_{\rm inv}$, $S_3$ is predicted to be much below unity. Neither the $^3_\Lambda$H wave function uncertainty, nor the details of working with a 1D vs. 3D HBT source parameterisation are expected to enable, e.g., $S_3>0.2$ at $R_{\rm inv}\approx1$~fm, characteristic for pp collisions. 

\section{Discussion and summary.}\label{s:conc}
The formation of nuclei by coalescence and femtoscopic [or Hanbury Brown-Twiss (HBT)] correlations between continuum proton pairs are two manifestations of final state interactions (FSI) that ``dress" an underlying high-excitation state (HXS) produced in hadronic collisions. In Sec.~\ref{s:basics}, starting from a relativistic quantum field theoretic (QFT) computation, we recalled how the effective quantum mechanical (QM) description of the dynamics of pairs that are nonrelativistic in the pair rest frame (PRF) gives rise to the usual HBT formalism, with which a two-particle source, characterising the HXS, can be measured. The same two-particle source enters the coalescence formula for deuteron formation. Much of our analysis was based on the formalism reviewed by Lednicky~\cite{Lednicky:2005tb}; our contribution was to reduce this formalism to observationally accessible language in the context of nuclear clusters. In both cases -- HBT and coalescence -- effective QM holds for ${\bf q^2\ll m^2}$, where ${\bf q}$ is the PRF momentum difference and $m$ is the nucleon mass; and up to corrections of order $\sim0.2\frac{\rm 1fm}{\bf x}$, where ${\bf x}$ is the average PRF characteristic distance between the nucleons in the HXS. Although it should not be trusted beyond $\mathcal{O}(10\%)$ accuracy\footnote{Making some interesting proposed $\mathcal{O}(10\%)$ tests potentially challenging~\cite{Mrowczynski:2019yrr,Bazak:2020wjn}.}, the QM limit is useful for deriving simple physically-transparent formulae and should be sufficient for ruling out (or supporting) coalescence as the origin of nuclei in hadronic collisions. In principle, while we did not pursue this route, the QM approximation can be avoided if one works directly with the relativistic FSI Bethe-Salpeter amplitudes. 


Coalescence after kinetic freeze out must occur at some level in hadronic collisions, but this does not mean that it is necessarily the dominant origin of clusters. The key assumption made by the coalescence model is that the long-range action of FSI can be factored out of an underlying short-range HXS dynamics. The level of accuracy of this factorisation is not obvious. To some extent, the apparent success of HBT analyses~\cite{Adam:2015vja,Acharya:2018gyz,Acharya:2020dfb} in reconstructing a fit of the two-nucleon source, using sophisticated physical FSI calculations, supports the coalescence framework. But we feel that this is not yet fully convincing: while the HBT analyses~\cite{Adam:2015vja,Acharya:2018gyz,Acharya:2020dfb} reported measurements of the two-nucleon source with a stated precision of $\mathcal{O}(10\%)$, they did so using a naive 1D Gaussian source fit\footnote{This is true also for~\cite{Acharya:2020dfb}, which assumed a 1D core source and modulated it by strong resonance decays.}. However, the true underlying source is not expected to be 1D; in fact, at least for mean lab frame momentum ${\bf p}\gtrsim m$, it must be anisotropic at the $\mathcal{O}(1)$ level -- as seen in the PRF -- due to Lorentz contraction. We can add to this the chaoticity parameter $\lambda_2$ (denoted just $\lambda$ in most of the literature), which was found to be significantly smaller than 1 in~\cite{Adam:2015vja}, but was held fixed to 1 in the fit of~\cite{Acharya:2018gyz,Acharya:2020dfb}. Altogether, it is clear that higher statistics HBT analyses have an important future role to play~\cite{Citron:2018lsq} in establishing to what extent true physical HXS information is revealed in femtoscopy. 
Our work highlights the importance of this question also to the origin of clusters.

Keeping the caveats above in mind, the main point of our work is to establish the coalescence/femtoscopy framework as a means to test coalescence by grounding the coalescence predictions with HBT information. 
The coalescence--correlations relation between HBT and deuteron formation, summarised by {\bf Eqs.~(\ref{eq:Cs},\ref{eq:B2})} [or equivalently {\bf Eqs.~(\ref{eq:C2mom},\ref{eq:B2mom})} in momentum space], was derived first in~\cite{Blum:2019suo} in the QM limit. The relation is very general and model-independent. As shown in~\cite{Blum:2019suo}, it is consistent with the early model-dependent derivation of~\cite{Scheibl:1998tk} (the latter, however, specialised to a particular model of hydrodynamical flow and was limited to small $p_t$). It is also consistent with the body of work by Mrowczynski~\cite{Mrowczynski:1987oid,Mrowczynski:1989jd,Mrowczynski:1992gc,Mrowczynski:1993cx,Mrowczynski:1994rn,Maj:2004tb,Mrowczynski:2016xqm}.

The connection of HBT with deuteron formation does not rely on density matrix factorisation. To derive formulae for three-body clusters, though, factorisation is needed. Assuming factorisation we derived the formulae for $^3_\Lambda$H and $^3$He, summarised in {\bf Eqs.~(\ref{eq:B3H},\ref{eq:B3})} [or {\bf Eqs.~(\ref{eq:B3Hmom},\ref{eq:B3mom})} in momentum space]. As an aside, we note that the derivation does not leave room for the confusion between ``pn$\Lambda$ channel" and ``D$\Lambda$ channel" advocated in~\cite{Zhang:2018euf,Sun:2018mqq} for $^3_\Lambda$H. Three-body coalescence comes from three-body FSI and is captured by a single formula, involving the three-body nucleus wave function.

In Sec.~\ref{s:hbtB} we combined formulae for the two-nucleon source, constrained by experimental HBT fits, with nuclear wave functions to obtain expressions for coalescence factors that can be compared to data. Exploring different forms for the wave function of $^3_\Lambda$H, we showed that although a full numerical representation of the wave function deviates significantly from a Gaussian, the numerical impact on the coalescence factor is modest, about a factor of two (with the more realistic numerical wave function predicting a higher coalescence factor). Our calculations allow for different scales in the wave functions of three-body states. This is particularly important for  $^3_\Lambda$H where the pn-$\Lambda$ factor is significantly more extended than the effective pn factor. Some earlier implementations~\cite{Bellini:2018epz} of the coalescence--correlations relation for $^3_\Lambda$H did not account for this fact.  

Our calculations also account for the anisotropic shape of the two- and three-nucleon source describing the HXS. As we illustrate in App.~\ref{sss:model}, the two-particle source is expected to be truly anisotropic in nature, especially at large $p_t$. 

In Sec.~\ref{s:dat} we compared our theoretical predictions to data. {\bf Fig.~\ref{fig:BR}} shows that coalescence, calibrated by HBT, is  consistent at the $\mathcal{O}(1)$ level with the $p_t$-differential yields of D and $^3$He for systems ranging from pp, pPb to PbPb and across different centralities. This comparison spans a dynamical range of $\sim30$ for $\B_2$ and $\sim10^{3}$ for $\B_3$. Some tension, at the $\sim2\sigma$ level or so, is seen for pp in both of the D and $^3$He yields. 
This situation gives strong motivation for a dedicated experimental analysis, studying HBT and cluster yields side by side in the same data set under the same kinematic conventions and cuts. On the HBT side, we urge the experiments to report HBT measurements allowing the chaoticity parameter $\lambda_2$ to float in the fit. In addition, as much as statistics permits, measurements of the 3D source (in the out-side-long parameterisation) would be preferred over 1D measurements of $R_{\rm inv}$. Such combined experimental analysis could zoom in on the coalescence--correlations prediction beyond the $\mathcal{O}(1)$ level by which it can currently be tested. This could, in principle, sharpen tensions with D and/or $^3$He data where they are currently difficult to establish conclusively.

Hypertriton $^3_\Lambda$H is known as a sensitive test of coalescence because of its large size, suggesting that the QM factor discriminating coalescence from the statistical hadronisation model (SHM) should be small and discernible. Our calculation, depicted in {\bf Fig.~\ref{fig:B3L}}, shows that the current measurements of $\B_{3\Lambda}$ in PbPb collisions are consistent with the coalescence prediction. 
The observable $S_3=\B_{3\Lambda}/\B_3$ may be more robust than $\B_{3\Lambda}$, because some experimental and theoretical uncertainties may cancel in the ratio. We compare our calculation of $S_3$ with the data in {\bf Fig.~\ref{fig:S3}}, finding some tension: the $S_3$ data tends to be somewhat higher than coalescence predicts. This discrepancy is not (yet) very significant, around $2\sigma$. A higher statistics measurement, and measurements in small systems like pp, pPb, or low multiplicity PbPb collisions -- in short, systems with small HBT radius -- would provide a critical test of coalescence.

\acknowledgments
We thank Fabian Hildenbrand for providing us with numerical calculations of the $^3_\Lambda$H wave function, the Fabbietti TUM group, Urs Wiedemann and the participants of the CERN workshop ``Origin of nuclear clusters" for insightful discussions and Nitsan Bar and Yossi Nir for comments on the manuscript. 
KB is incumbent of the Dewey David Stone and Harry Levine career development chair at the Weizmann Institute of Science.

\begin{appendix}

\section{Corrections to the Gaussian source approximation.}\label{ss:SnonG}
The Gaussian source considered in Sec.~\ref{ss:GS} cannot be exact.
It is therefore important to estimate the uncertainty in the coalescence calculation, that comes about if this approximation is used. In this section we consider mechanisms that violate the Gaussian source approximation and estimate their quantitative impact.

\subsection{Feed-down from weak decays and source chaoticity.}\label{sss:weak}
In comparing theory to data one should account for the effect of feed-down from weak decay, where a particle that is emitted from the HXS as $\Lambda$ or $\Sigma$ decays into a proton in the detector (see~\cite{Wiedemann:1996ig} for a parallel discussion for pions). The decay vertex is displaced by $\mathcal{O}(1~{\rm cm})$ from the HXS, which means that the FSI relevant for two-particle correlation are those involving $\Lambda$ or $\Sigma$. 
Ref.~\cite{Acharya:2020dfb} estimated that a fraction $\alpha_{p}\approx0.82$ of detected protons in their analysis originate from a genuine emitted proton, while a fraction $\alpha_{\Lambda}\approx0.12$ and a fraction $\alpha_{\Sigma}\approx0.06$ of detected protons originate from an emitted $\Lambda$ or $\Sigma$, respectively\footnote{The feed-down fractions in~\cite{Acharya:2020dfb} were allowed to vary by 20\% as part of the systematic uncertainty estimate. Indeed, these fractions were different by $\sim20\%$ in Ref.~\cite{Acharya:2018gyz}, although this may be in part due to different experimental kinematical cuts and selection criteria.}. Neglecting particle misidentification (which add up to about 1\% in~\cite{Acharya:2020dfb}) we have $\alpha_{p}\approx1-\alpha_{\Lambda}-\alpha_{\Sigma}$. 
Thus, {\it if HBT correlations are ignored}, a fraction $\lambda^{(pp)}\approx(1-\alpha_{\Lambda}-\alpha_{\Sigma})^2\approx0.7$ of detected pp pairs come from genuine emitted pp, while, for example, a fraction $\lambda^{(p\Lambda)}\approx2\alpha_{\Lambda}(1-\alpha_{\Lambda}-\alpha_{\Sigma})\approx0.2$ of detected pp pairs come from an emitted p$\Lambda$ pair, etc. 

In constructing the observable two-proton correlation, Refs.~\cite{Adam:2015vja,Acharya:2018gyz,Acharya:2020dfb} divided the measured proton pair spectrum by the spectrum of pairs from uncorrelated mixed events, without feed-down subtraction. 
Then, in fitting a model of the correlation to the data, the correlation function was split into a part coming from genuine emitted pp pairs and parts coming from emitted p$\Lambda,\,\Lambda\Lambda,\,p\Sigma$ and $\Sigma\Sigma$ pairs, using the relevant FSI for each part\footnote{Ref.~\cite{Niedziela:2018fau} also used a similar formalism in studying baryon--anti-baryon correlations.}:

\be\label{eq:2pmod} C^{\rm model}({\bf p,q})&=&1+\sum_{i={\rm pp,p\Lambda,...}}\lambda^{(i)}\left[C^{(i)}({\bf p,q})-1\right],
\ee
 with the channel-specific weights and correlation functions
\be \label{eq:Cmod}C^{(i)}({\bf p,q})&=&\sum_sw_s\int d^3{\bf r}\,\mathcal{S}_2^{(i)}({\bf r})\left|\phi^{(i)}_{s,q}({\bf r})\right|^2.\ee
Here, for example, for $i=p\Lambda$ the wave function at spin channel $s$ is given by $\phi^{(p\Lambda)}_{s,q}({\bf r})$, etc. In Ref.~\cite{Acharya:2020dfb}, the nucleon source functions $\mathcal{S}_2^{(i)}({\bf r})$ were modelled differently for different $i$, to account for strong resonances that have a different characteristic decay range for p and $\Lambda$ daughters. In contrast, Refs.~\cite{Adam:2015vja,Acharya:2018gyz} assumed a common nucleon source $\mathcal{S}_2^{(i)}=\mathcal{S}_2$.

As long as the $\lambda^{(i)}$ parameters are correctly calibrated to account for weak decays, a quick check verifies that the correlation functions $C^{(i)}$ in Eq.~(\ref{eq:Cmod}) do indeed match with the theoretical definition of the same objects as derived in Sec.~\ref{ss:2ptD}. Refs.~\cite{Acharya:2018gyz,Acharya:2020dfb} assumed that this was the case and fixed the numerical values of the $\lambda^{(i)}$ according to the experimentally determined single proton purity. 
In contrast, Ref.~\cite{Adam:2015vja} did not fix the value of $\lambda^{(pp)}$ and $\lambda^{(p\Lambda)}$, but rather considered these parameters as part of the experimental fit. Interestingly, the fit resulted in $\lambda\equiv\lambda^{(pp)}+\lambda^{(p\Lambda)}\approx0.3-0.7$, significantly less than unity\footnote{Useful details can be found in~\cite{Szymanski:2016xia}. For comparison, the single-particle purity estimates of~\cite{Acharya:2018gyz,Acharya:2020dfb} read $\lambda^{(pp)}+\lambda^{(p\Lambda)}\approx0.87$ and $0.9$, respectively; the remaining probability being associated mostly with p$\Sigma$ pairs.}.

From the theoretical point of view, the approach of~\cite{Adam:2015vja} is beneficial over that of~\cite{Acharya:2018gyz,Acharya:2020dfb}. The reason has to do with the modelling of the source function $\mathcal{S}_2({\bf r})$~\cite{Heinz:1999rw,Blum:2019suo}. 
Refs.~\cite{Adam:2015vja,Acharya:2018gyz,Acharya:2020dfb} assumed in their fit the 1D Gaussian form Eq.~(\ref{eq:S_2(r)isoGauss}). This source function integrates to unity, $\int d^3{\bf r}\,\mathcal{S}_2({\bf r})=1$, corresponding to $\C_2({\bf q}=0)=1$. However, there is no a-priori reason to assume that the true $\C_2$ (or $\mathcal{S}_2$) satisfy this normalisation exactly. In fact, observing a departure from this normalisation in the data could hint, for example, at a violation of the factorisation assumption of Sec.~\ref{ss:factorisation}. In addition, it is well known that if the experimental fit assumes a Gaussian form for $\C_2({\bf q})$, but the true  $\C_2({\bf q})$ is non-Gaussian, then adding an intercept parameter $\lambda_2$ to the fit, as in Eq.~(\ref{eq:C2ssng}), can absorb some of the difference. 

We can thus interpret the $\lambda$ measurement of~\cite{Adam:2015vja} as a measurement of the parameter $\lambda_2$, via $\lambda_2\approx\lambda$. In contrast, the analysis of Refs.~\cite{Acharya:2018gyz,Acharya:2020dfb} effectively forced $\lambda_2\to1$ in the fit procedure.

Considering now the coalescence factor, a direct comparison to the HBT analysis is possible if the single proton spectrum used in the experimental definition of $\B_A$ (e.g. Eq.~(\ref{eq:B20})) is obtained subtracting the weak decay feed-down contributions. With this the coalescence--correlation relation of Eq.~(\ref{eq:B2}) or Eq.~(\ref{eq:B2mom}) applies as is. 
For $\B_{3,3\Lambda}$, a fully data-driven analysis would require an experimental measurement of $\C_3$. In the absence of that, we can roughly estimate that for $\lambda_2\neq1$ the Gaussian source expressions for $\B_{3,3\Lambda}$ should be modified by $\B_{3,3\Lambda}\to\lambda_2^{\frac{3}{2}}\B_{3,3\Lambda}$~\cite{Blum:2019suo}. 

Unrelated to weak decays, a correction to the expressions of Sec.~\ref{s:hbtB} arises from the fact that Ref.~\cite{Acharya:2020dfb} actually found different fit results for the sources $\mathcal{S}_2({\bf r})$ deduced from pp and p$\Lambda$ pairs. One physically motivated reason for the difference stems from the different decay range of the strong resonances, that feed into p and $\Lambda$ states in the HXS. The same effect also predicts a non-Gaussian form for $\mathcal{S}_2({\bf r})$ (and $\C_2({\bf q})$), as we discuss next.

\subsection{Strong resonance decays.}\label{sss:strongres}
\begin{figure}[htbp]
  \begin{center}
   \includegraphics[width=0.495\textwidth]{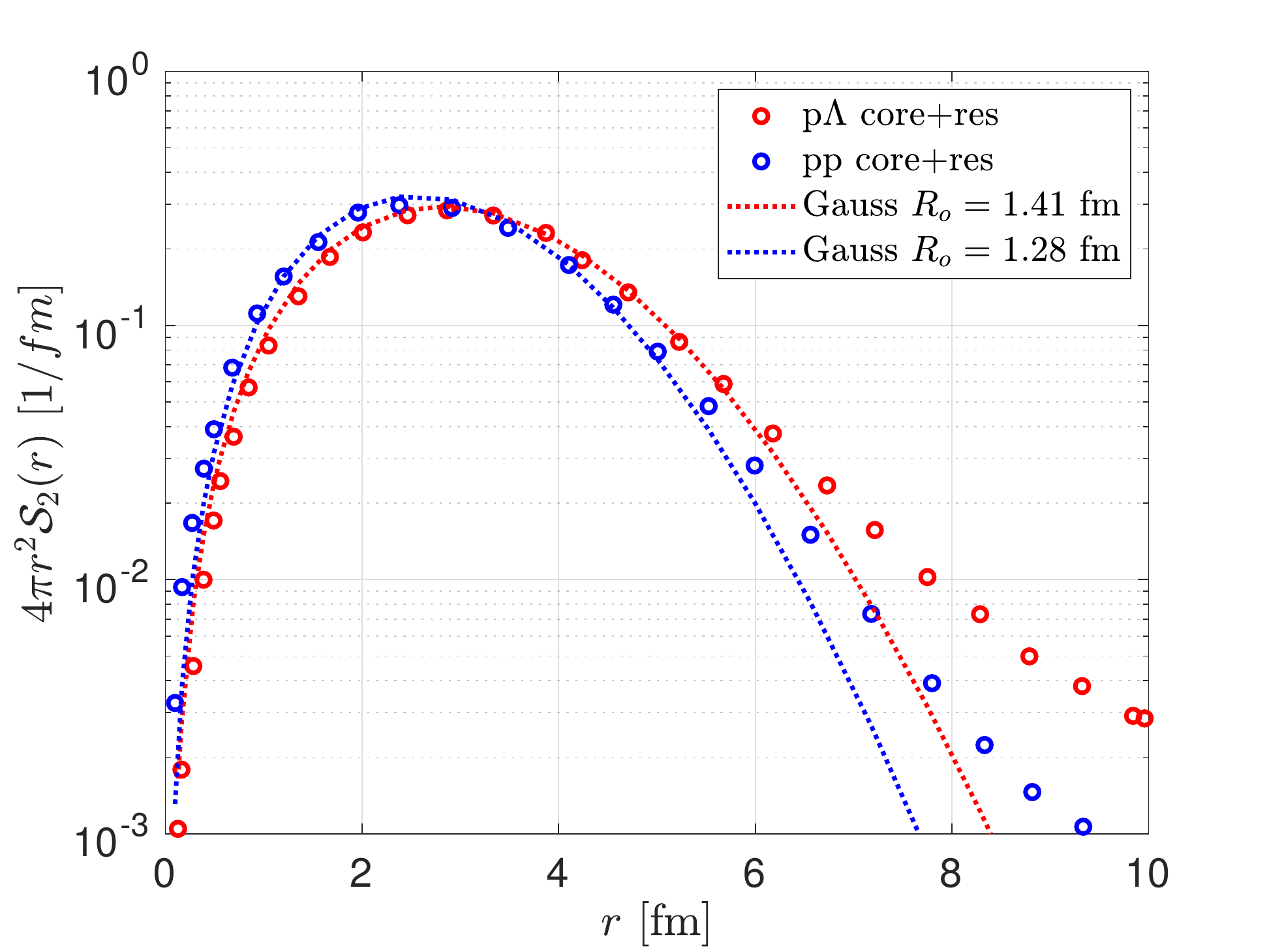}
   \includegraphics[width=0.495\textwidth]{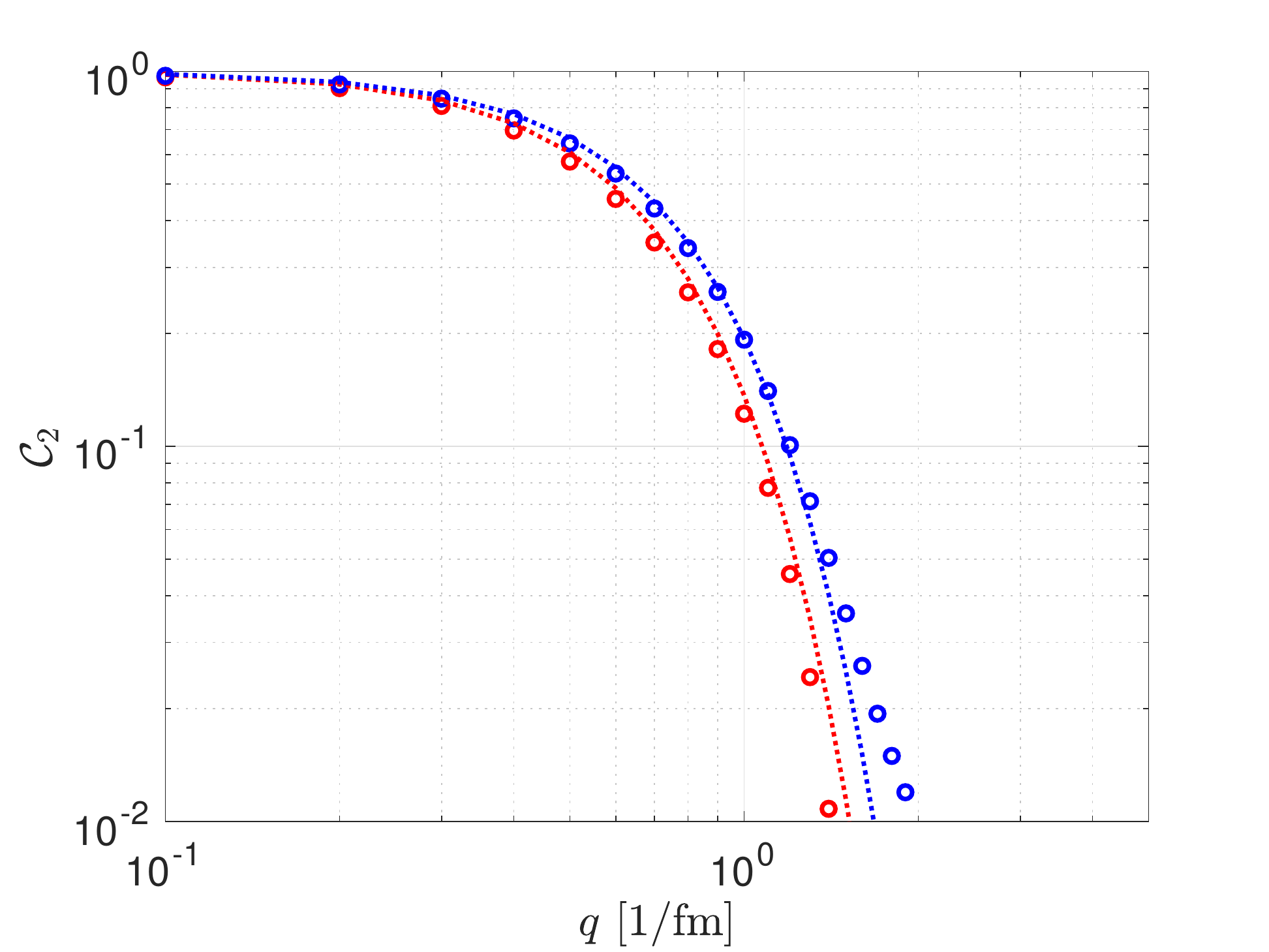}
  \end{center}
  \caption{{\bf Left:} The source $\mathcal{S}_2(\bf r)$, reproduced from the experimental fit result of Fig.~4 in Ref.~\cite{Acharya:2020dfb}. {\bf Right:} $\C_2$, the Fourier transform of the sources on the left.}
  \label{fig:SnonG}
\end{figure}

Even if the underlying emission function was an exact Gaussian, the decay of strong resonances with lifetimes of the order of a few fm would distort the effective source. 
Ref.~\cite{Acharya:2020dfb} studied this problem for pp and p$\Lambda$ correlations in pp collisions. An isotropic source model including strong resonance decay, which was found in~\cite{Acharya:2020dfb} to give an adequate fit to the correlation data, is reproduced here in the left panel of Fig.~\ref{fig:SnonG}. The full (non Gaussian) source is shown by circles, compared with the Gaussian source in dotted lines. Strong resonances lead to a non Gaussian tail of $\mathcal{S}_2({\bf r})$. 
In the right panel of Fig.~\ref{fig:SnonG} we show the $\C_2({\bf q})$ curves corresponding to the $\mathcal{S}_2({\bf r})$ curves on the left. 

Quantitatively, despite the strong resonance contribution, the Gaussian approximation can be seen to give a rather accurate description of the source: the difference between the full non Gaussian $\C_2$ and the Gaussian approximation is smaller than $\sim10\%$ throughout the range where $\C_2>0.1$. While Ref.~\cite{Acharya:2020dfb} dealt with pp initial state, we might expect that in larger systems like PbPb the relative impact of strong resonance decay could be even less significant. 

We emphasise that Ref.~\cite{Acharya:2020dfb}, which provided the example for our discussion, {\it assumed} an underlying Gaussian source which they then deformed by strong resonances. What we learn from this exercise, then, is that if the underlying ``genuine" emission function is Gaussian, then the effective two-particle source after strong decays is also consistent with Gaussian to 10\% accuracy, in the range of ${\bf r}$ (or, in momentum space, ${\bf q}$) that is relevant for coalescence. 
We stress, however, that while the assumed 1D Gaussian source of~\cite{Acharya:2020dfb} was experimentally consistent with HBT data, there is no guarantee that the true physical source is Gaussian. In App.~\ref{sss:model} we recall theoretical reasons to expect otherwise.

\section{Deviations from an isotropic Gaussian source in the phenomenological blast wave model.}\label{sss:model}
Relativistic expansion of the HXS ``fire ball" is expected to proceed differently along and transverse to the beam line. This unisotropic flow predicts that the source $\mathcal{S}_2({\bf r})$ should depend on the direction of ${\bf r}$ w.r.t. the beam line and also w.r.t. the pair mean momentum vector ${\bf p}$. Even if the nucleon source was somehow isotropic in the lab frame, it would not be seen as isotropic in the PRF for ${\bf p}\neq0$, due to Lorentz contraction. 
These effects were not modelled in the experimental analyses of Refs.~\cite{Adam:2015vja,Acharya:2018gyz,Acharya:2020dfb}, that performed fits to the 1D source Eqs.~(\ref{eq:C2ss}). Here we consider these effects from the point of view of the  phenomenological blast wave model (BWM). Our goal is not to argue either in favour or against the validity of the model, but simply to gain some feeling for the possible systematic error associated with adopting the 1D Gaussian source approximation, if that is applied to coalescence calculations via the coalescence--correlation relation.

To recall the BWM, we take the beam line to be along the $\hat z$ axis. The space-time coordinates are chosen as $\tau=\sqrt{t^2-z^2}$, $\rho=\sqrt{x^2+y^2}$, $\eta={\rm arctanh}(z/t)$, and the azimuthal angle $\phi$. The position 4-vector is then $R^\mu=\left(\tau\cosh\eta,\rho\cos\phi,\rho\sin\phi,\tau\sinh\eta\right)$ and $d^4R=d\tau \tau \,d\rho \rho\, d\eta \,d\phi$. The 4-momentum vector of a particle emitted at rapidity $Y$, azimuthal angle $\Phi$, and transverse momentum $p_t$, is $p^\mu=\left(m_t\cosh Y,p_t\cos\Phi,p_t\sinh\Phi,m_t\sinh Y\right)$, with $m_t=\sqrt{m^2+p_t^2}$. 
The single-particle emission function is parameterized using 6 parameters, $\tau_0,\Delta\tau,R_0,\beta_S,n,T$ (we neglect chemical potentials), as follows ($\theta(x)$ is the Heaviside step function):
\be \tilde S_p(R)&=&\sqrt{\frac{2}{\pi}}\frac{m_t}{m}\cosh\left(\eta-Y\right)J(\tau)\theta\left(R_0-\rho\right)e^{-\frac{pu}{T}},\\
u^\mu(R)&=&\left(\cosh\eta\cosh\eta_t,\sinh\eta_t\cos\phi,\sinh\eta_t\sin\phi,\sinh\eta\cosh\eta_t\right),\\
\eta_t(\rho)&=&{\rm tanh}^{-1}\left(\frac{\rho^n}{R_0^n}\,\beta_S\right),\\
J(\tau)&=&\frac{1}{\sqrt{2\pi}\Delta\tau}e^{-\frac{(\tau-\tau_0)^2}{2\Delta\tau^2}}.
\ee
With these definitions we can evaluate $\C_2(p,{\bf q})$ numerically via Eq.~(\ref{eq:C2model}). In doing so, recall that we require $q=(0,{\bf q})$ specified in the PRF, while the emission function $\tilde S_p(x)$ depends on space-time coordinates in the lab frame. Defining $b=p/m$, we can write $q'$ in the lab frame as $q'^{0}={\bf bq},\;\;{\bf q}'={\bf q}+\frac{\bf bq}{1+b^0}{\bf b}$~\cite{Scheibl:1998tk}. 
For simplicity, in the examples below we choose $p=(m_t,p_t,0,0)$. With this choice $pu=m_t\cosh\eta\cosh\eta_t-p_t\sinh\eta_t\cos\phi$ and using the $q_l,q_o,q_s$ decomposition we have $qx=\frac{p_t}{m}q_o \tau\cosh\eta-\frac{m_t}{m}q_o\rho\cos\phi-q_s\rho\sin\phi-q_l\tau\sinh\eta$. 
We consider separately the cases ${\bf q}=(q_o,0,0)$, ${\bf q}=(0,q_s,0)$, and ${\bf q}=(0,0,q_l)$. 

\begin{figure}[htbp]
  \begin{center}
   \includegraphics[width=0.45\textwidth]{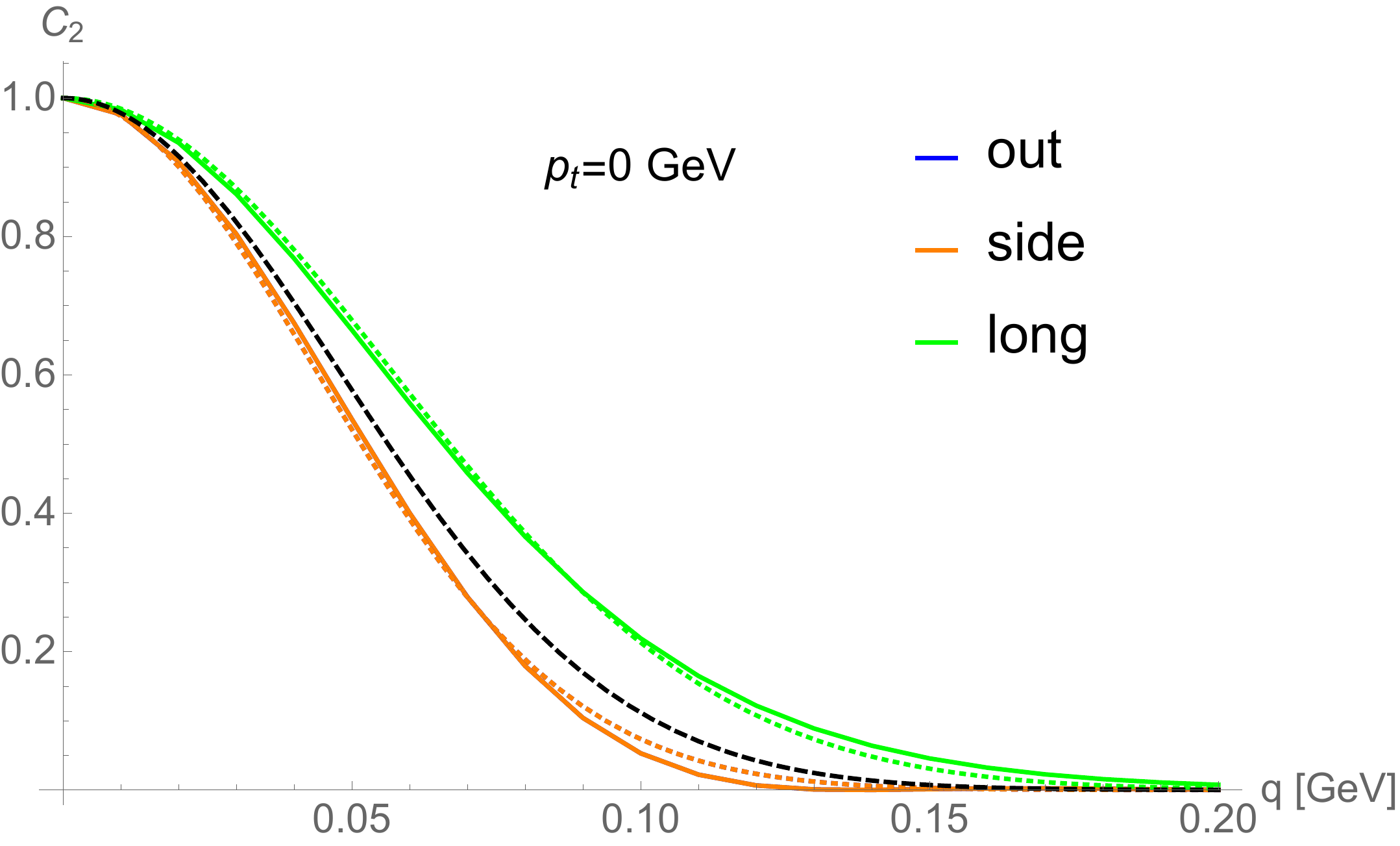}
      \includegraphics[width=0.45\textwidth]{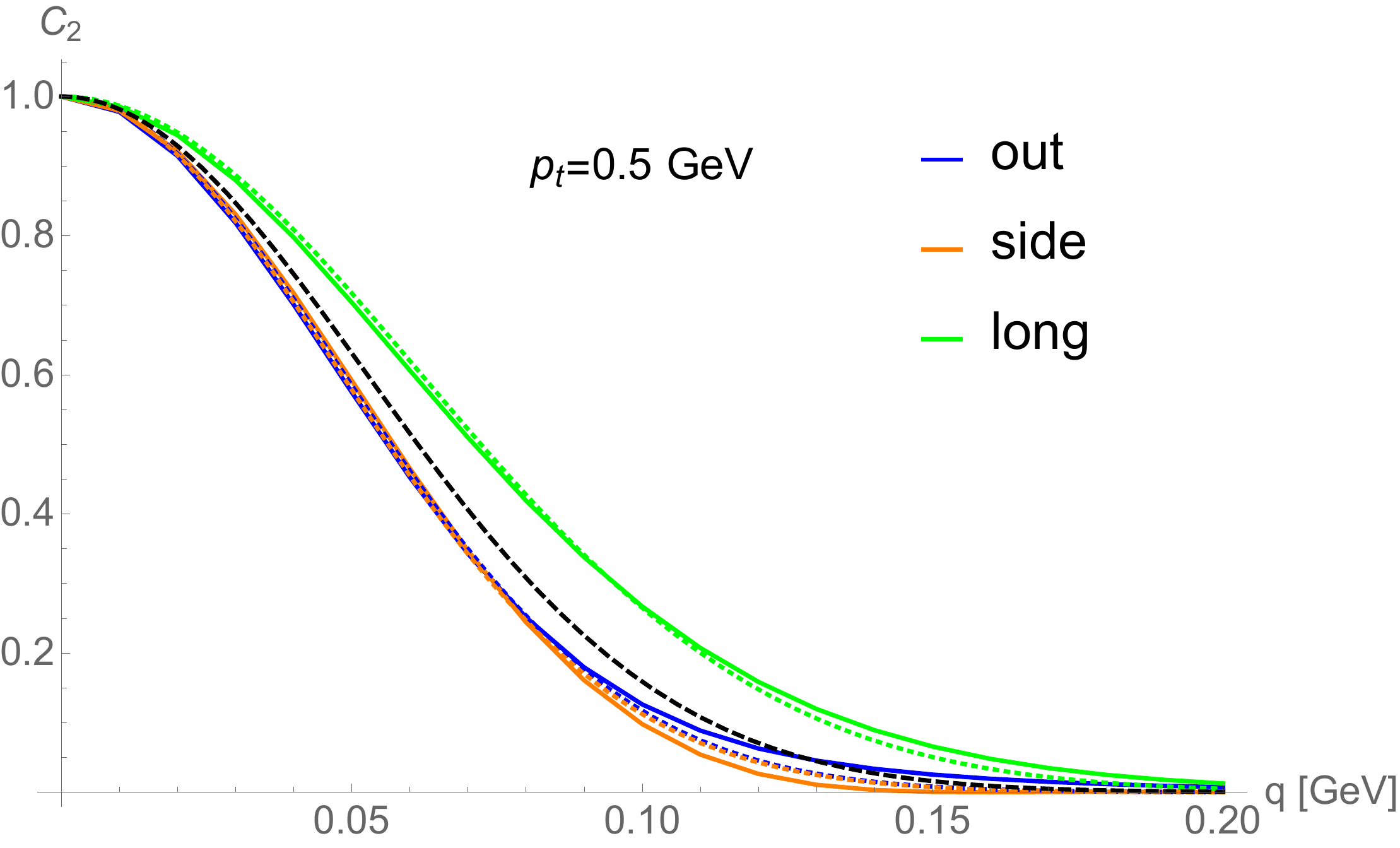}
      \includegraphics[width=0.45\textwidth]{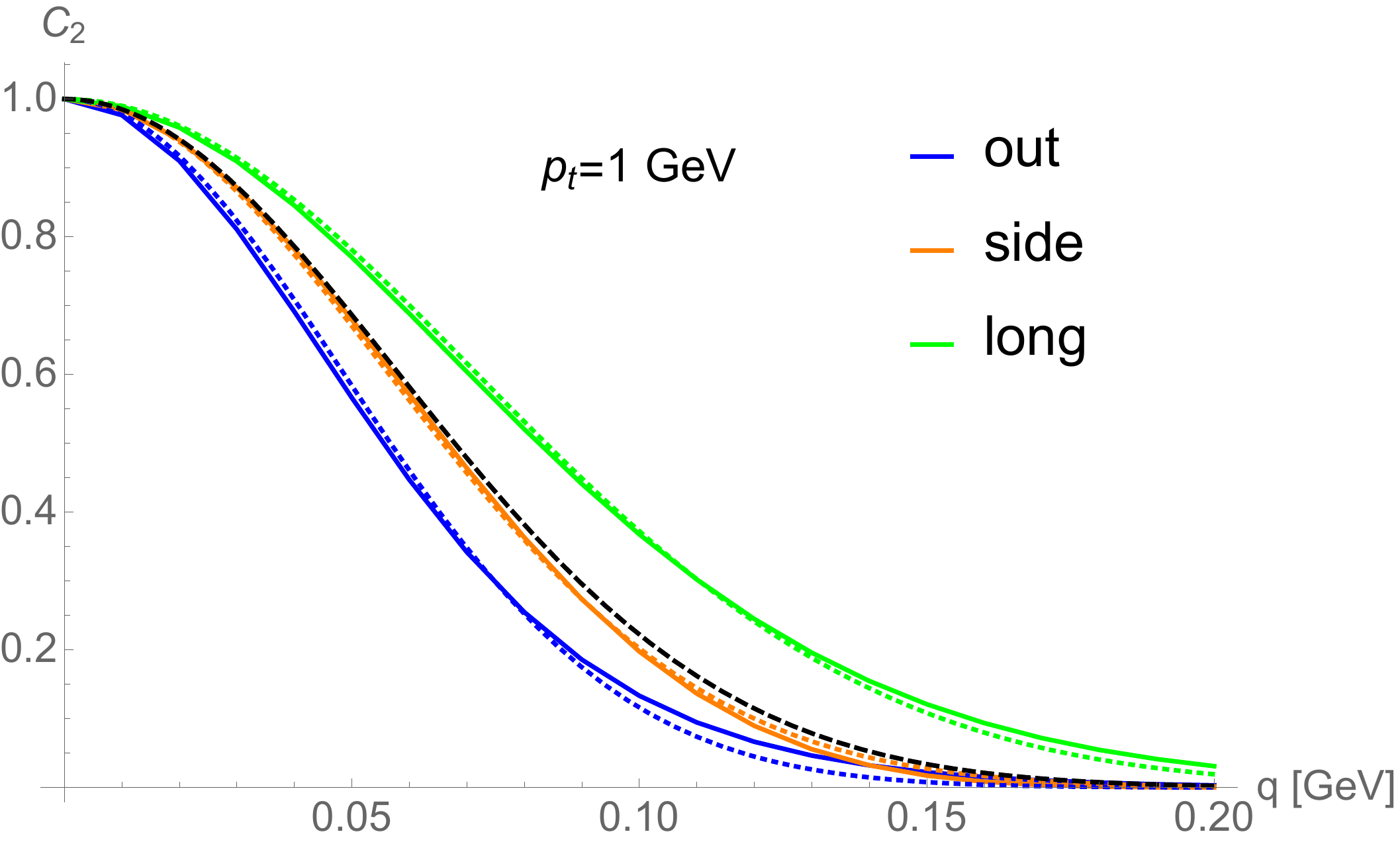}
         \includegraphics[width=0.45\textwidth]{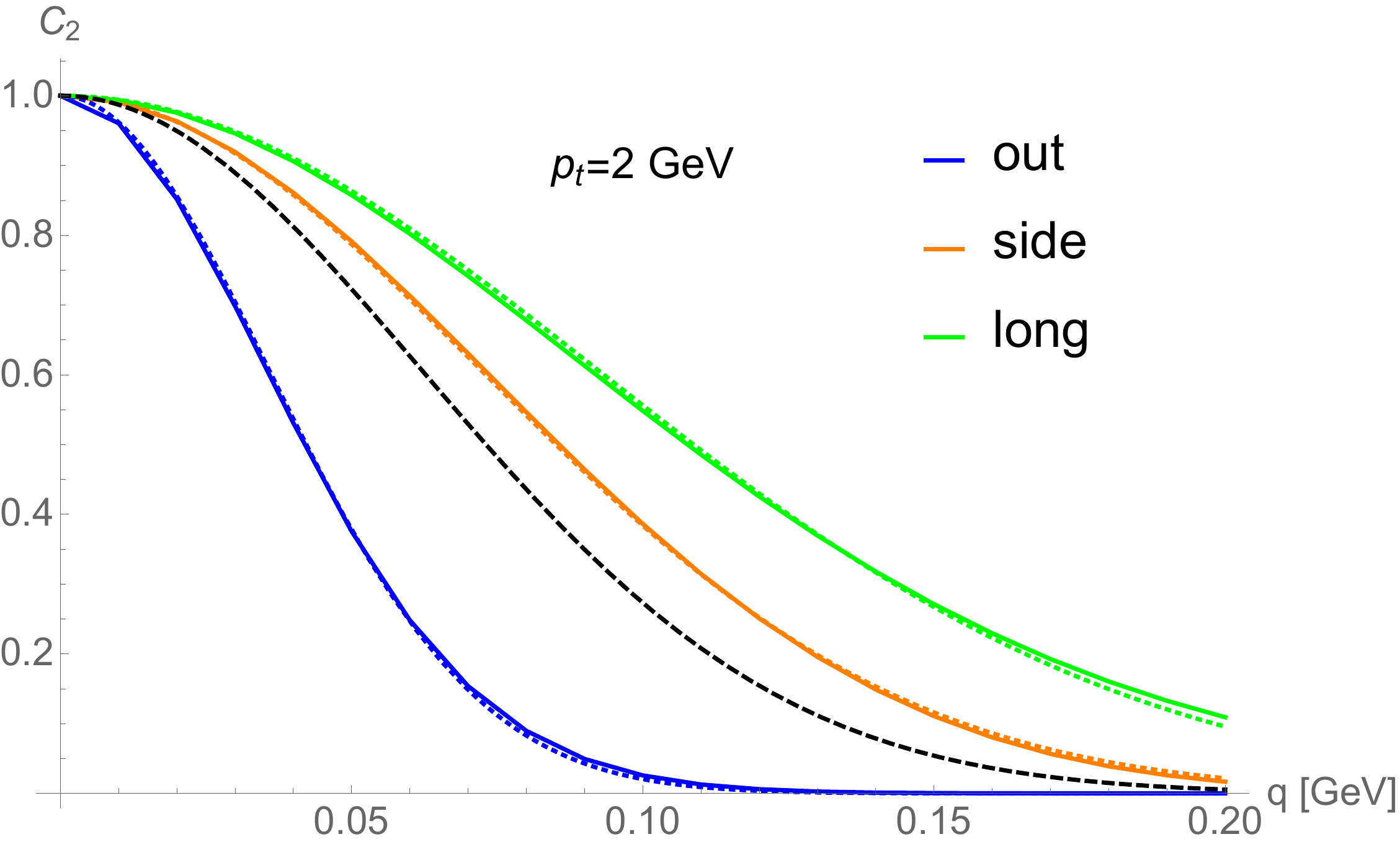}
  \end{center}
  \caption{The two-particle source $\C_2(p,{\bf q})$ calculated in the blast wave model and projected onto the out, side, long directions. Solid lines show the numerical result, while dotted lines show independent Gaussian fits to each projection. Black dashed line shows the 1D Gaussian source with $R_{\rm inv}=\left(R_oR_sR_l\right)^{\frac{1}{3}}$. The BWM parameters are chosen to mimic PbPb collisions at the LHC (see text). {\bf Top left, top right, bottom left, bottom right:} $p_t=0,0.5,1,2$~GeV.}
  \label{fig:C2BWM}
\end{figure}

In Fig.~\ref{fig:C2BWM} we plot $\C_2(p,{\bf q})$, projected onto the out, side, and long directions (solid blue, orange, and green, respectively). Independent Gaussian fits to each projection are shown by dotted lines. Black dashed line shows the 1D Gaussian source computed with $R_{\rm inv}=\left(R_oR_sR_l\right)^{\frac{1}{3}}$. For definiteness we take $T=100$~MeV, $n=1$, $\beta_s=0.5$ and $\tau_0=7$~fm, $\Delta\tau=1.5$~fm, $R_0=7$~fm, chosen to roughly represent PbPb collisions~\cite{Adam:2015vda}. Note that the phenomenological BWM is based on classical intuition and satisfies by construction -- effectively -- the density matrix factorisation assumption. Thus the chaoticity parameters $\lambda_2=\lambda_3=1$ in this computation, and we are guaranteed that $\C_2(p,{\bf 0})=1$.

\begin{figure}[htbp]
  \begin{center}
   \includegraphics[width=0.45\textwidth]{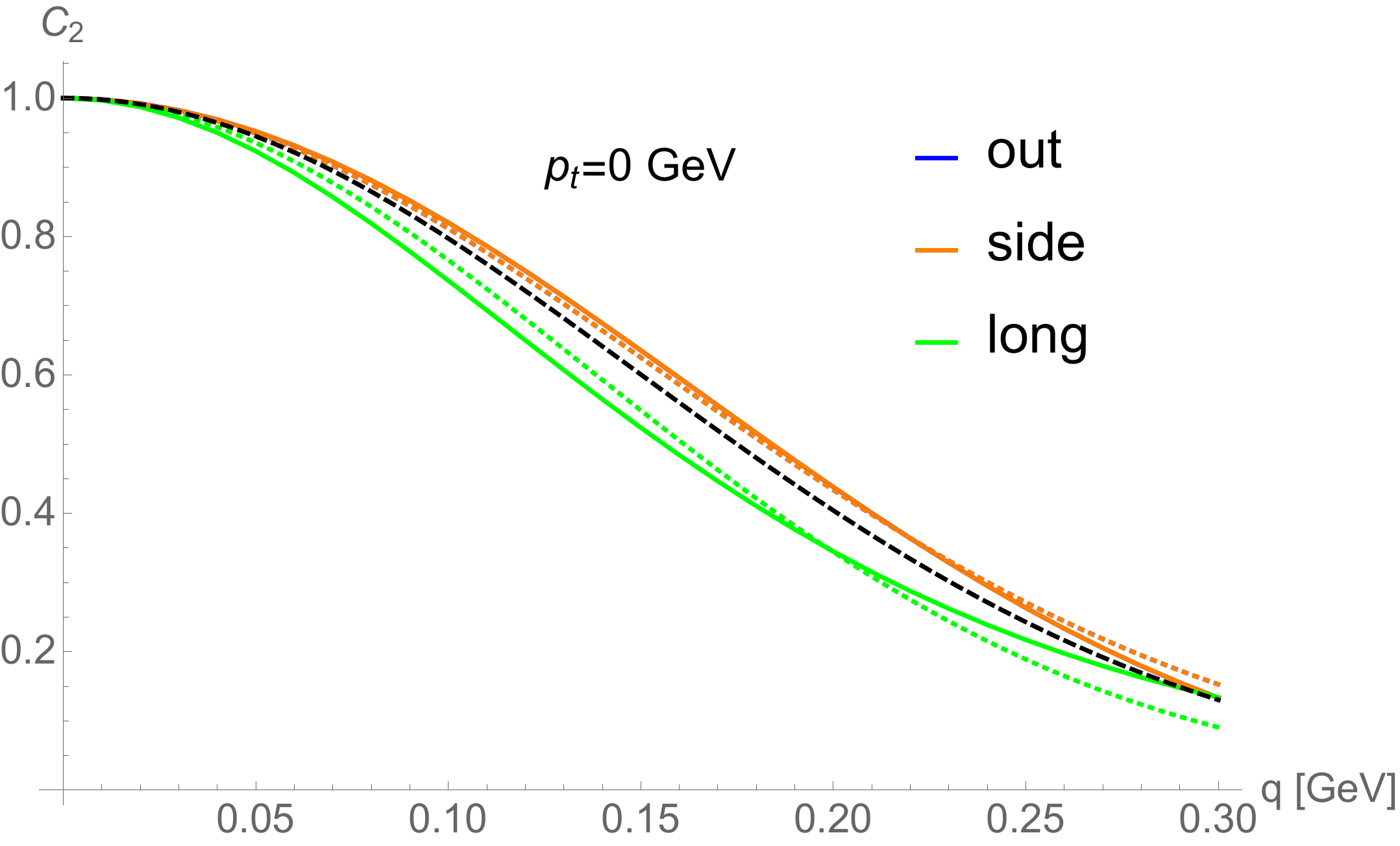}
      \includegraphics[width=0.45\textwidth]{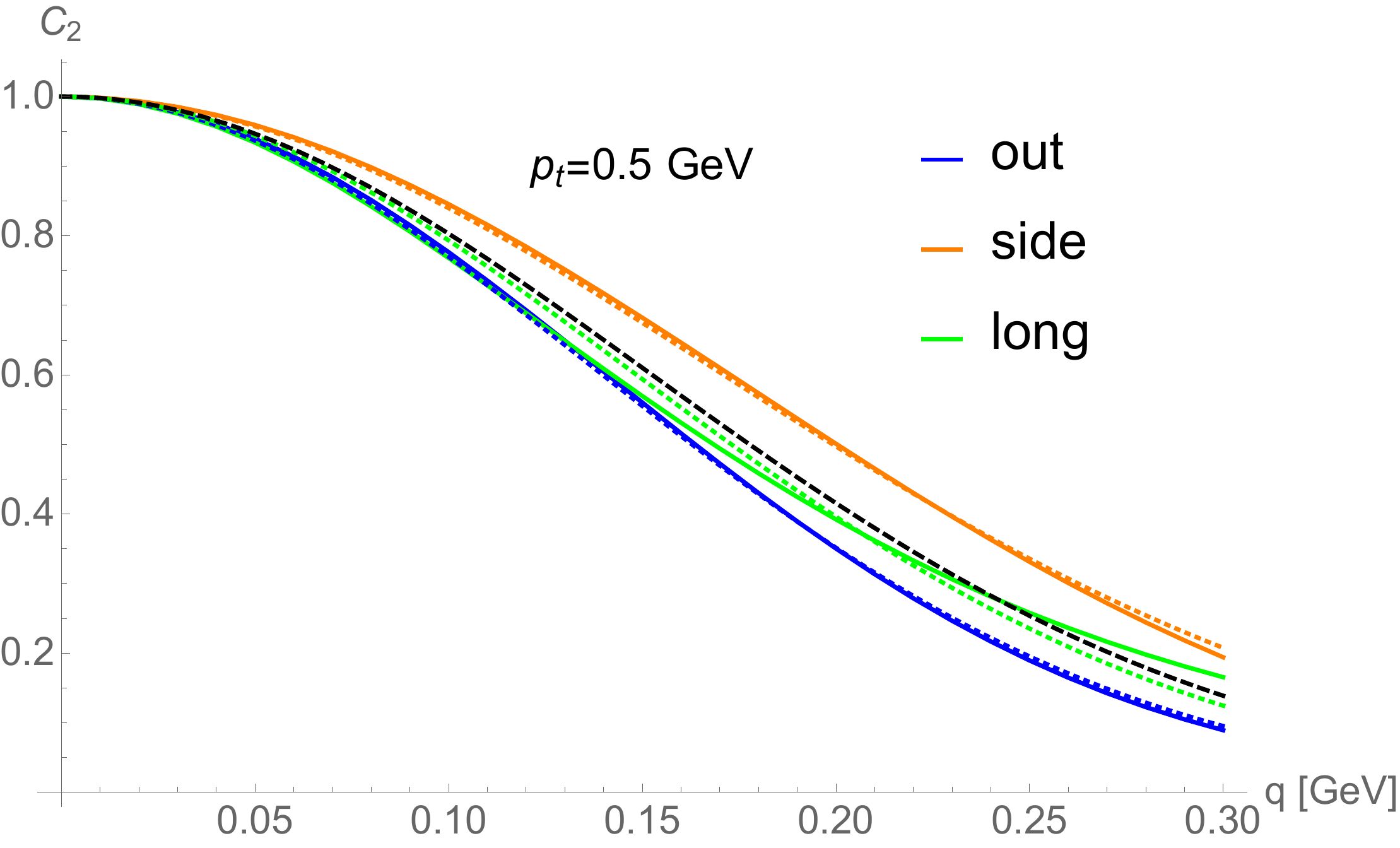}
       \includegraphics[width=0.45\textwidth]{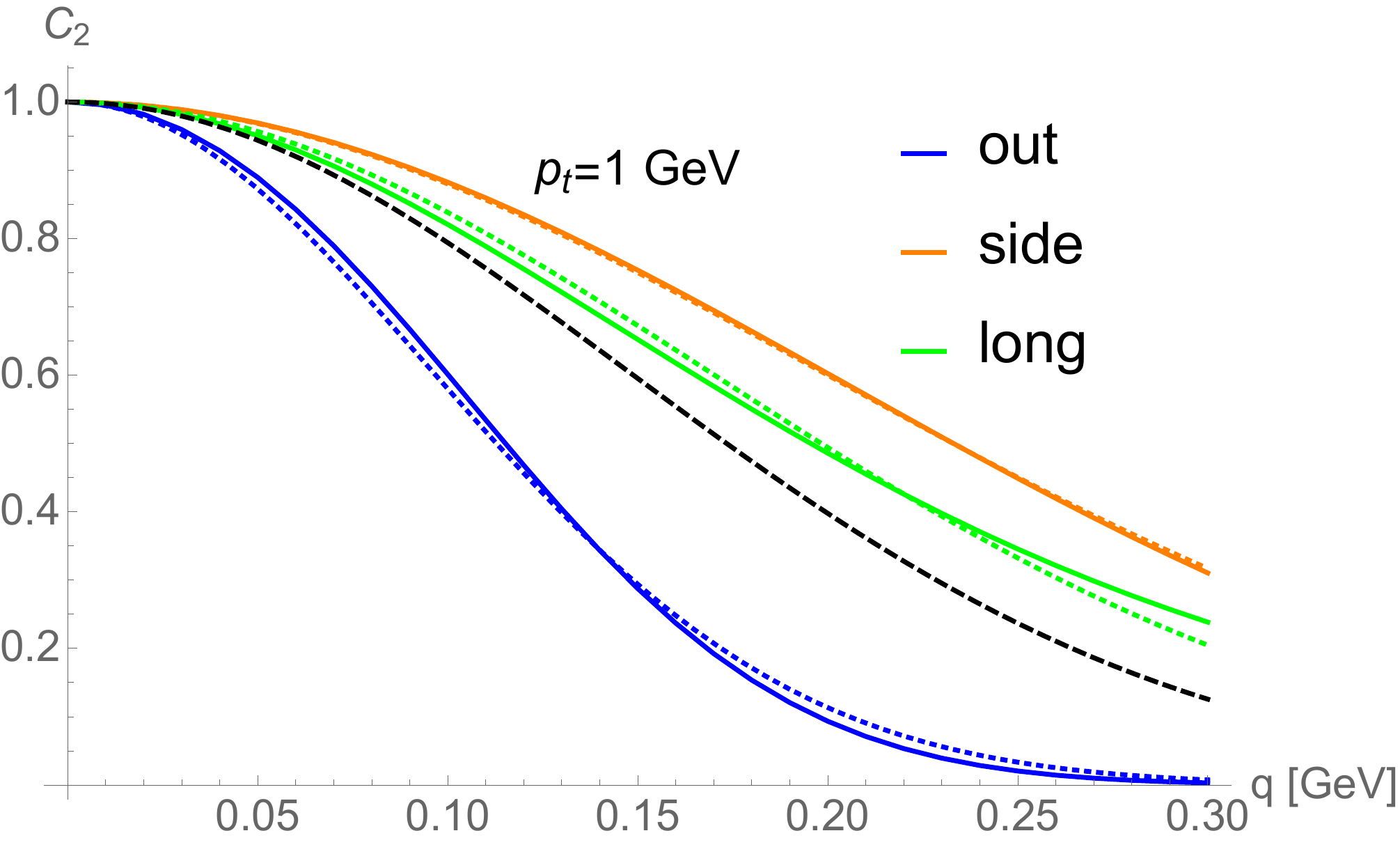}
         \includegraphics[width=0.45\textwidth]{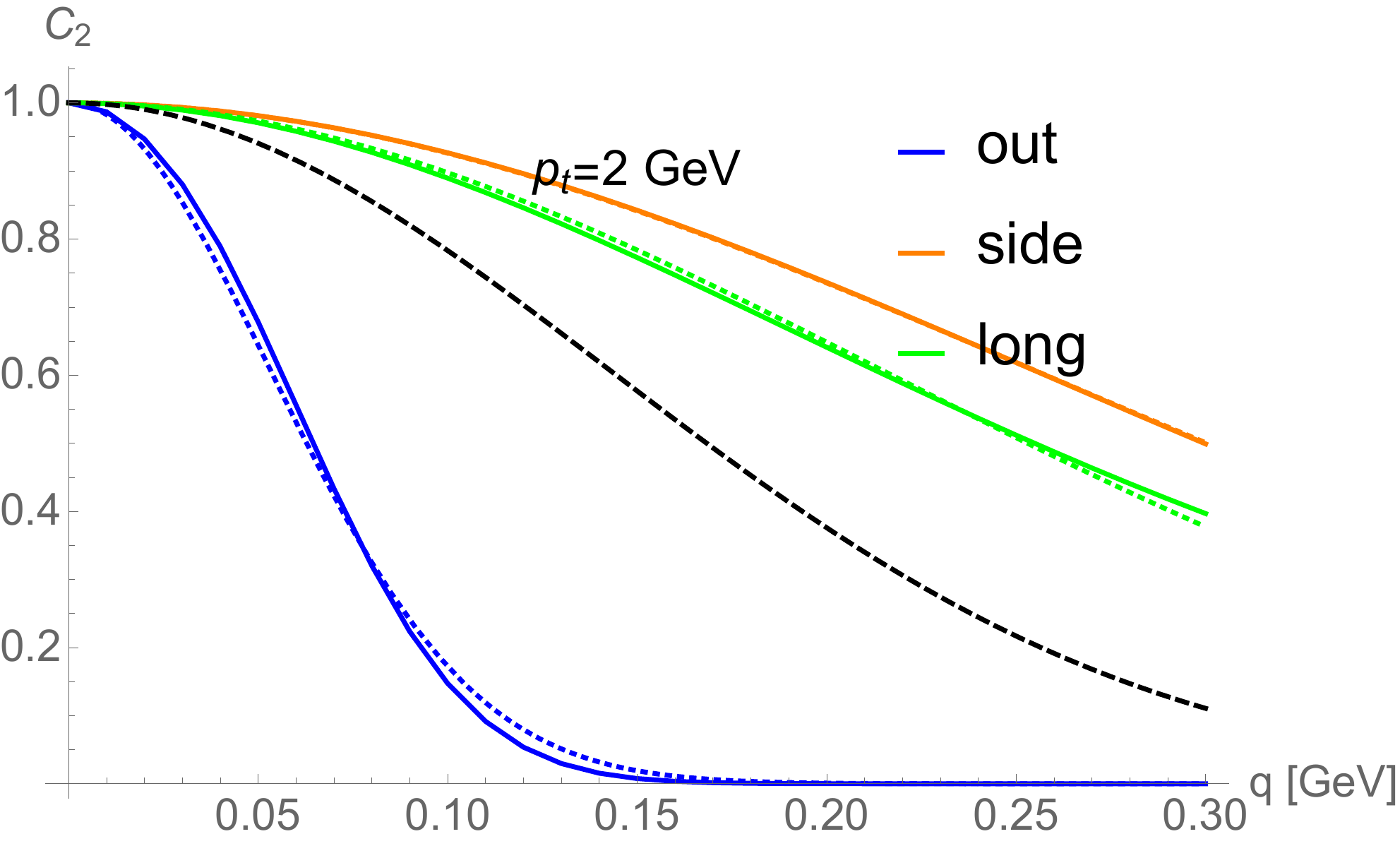}
  \end{center}
  \caption{Same as Fig.~\ref{fig:C2BWMpp}, with BWM model parameters chosen to mimic pp collisions at the LHC.}
  \label{fig:C2BWMpp}
\end{figure}

The source depicted in Fig.~\ref{fig:C2BWM} was chosen to resemble PbPb collisions. From the fits in the plot, the radii of homogeniety for this source at $p_t=0$, for example, are $R_o=R_s\approx3.2$~fm, $R_l\approx2.5$~fm, extended compared to the deuteron RMS radius $r_{rms}=\sqrt{3/8}b_d\approx2.1$~fm. To see the size of the effect for smaller systems, e.g. pp collisions, we recalculate $\C_2$ for different values of $R_0=\tau_0=2$~fm, leading to $R_o=R_s\approx0.9$~fm, $R_l\approx1$~fm at $p_t=0$. (The other BWM parameters are unchanged.) The results are shown in Fig.~\ref{fig:C2BWMpp}.

Figs.~\ref{fig:C2BWM}-\ref{fig:C2BWMpp} demonstrate how unisotropic flow leads to an unisotropic source. Notably, the width along the out direction shows Lorentz contraction in the PRF at $p_t>0$. 
This perspective calls into question the attempt in~\cite{Acharya:2020dfb} to fit the details of $\mathcal{S}_2({\bf r})$ to $\mathcal{O}(10\%)$ precision, based on the 1D isotropic Gaussian model.

Our main interest is to check to what extent the deviation from 1D Gaussian source  affects the coalescence calculation. Again from Figs.~\ref{fig:C2BWM}-\ref{fig:C2BWMpp} (as well as from the discussion in Sec.~\ref{sss:strongres}), we expect that a 3D Gaussian approximation may do a reasonably accurate job describing $\C_2$; thus we will use the 3D Gaussian parameterisation as a standard for comparison. For a rough estimate of the error, incurred by using the 1D fit of Eq.~(\ref{eq:C2ss}) to describe a 3D source, we calculate $R_o,R_s,R_l$ in the BWM and define 
\be R_{\rm inv}&=&\left(R_oR_sR_l\right)^{\frac{1}{3}}.\ee
We can now compare the results of using Eq.~(\ref{eq:B2G}) with $R_o=R_s=R_l\to R_{\rm inv}$, to the results of the original Eq.~(\ref{eq:B2G}):
\be\label{eq:B21D23D}\frac{\B_2^{1D}}{\B_2^{3D}}&\sim&\sqrt{\frac{\left(b_d^2+4R_l^2\right)\left(b_d^2+4R_o^2\right)\left(b_d^2+4R_s^2\right)}{\left(b_d^2+4R_{\rm inv}^2\right)^{3}}}.\ee
Note that in both limits $b_d\to0$ and $b_d\to\infty$ this correction factor is equal to 1.
The results are summarised in Tab.~\ref{tab:B2corr}. 

We give a rough estimate of the correction to $\B_{3\Lambda}$ in a similar way:
\be\label{eq:B31D23D}\frac{\B_{3\Lambda}^{1D}}{\B_{3\Lambda}^{3D}}&\sim&\sqrt{\frac{\left(b_{pn}^2+2R_l^2\right)\left(b_{\Lambda}^2+2R_l^2\right)\left(b_{pn}^2+2R_o^2\right)\left(b_{\Lambda}^2+2R_o^2\right)\left(b_{pn}^2+2R_s^2\right)\left(b_{\Lambda}^2+2R_s^2\right)}
{\left(b_{pn}^2+2R_{\rm inv}^2\right)^3\left(b_{\Lambda}^2+2R{\rm inv}^2\right)^3}}.\ee
The calculation for $\B_3$ is the same up to $b_{pn}\to b_\Lambda\to b_{\rm ^3He}$. The results are also summarised in Tab.~\ref{tab:B2corr}.
\begin{table}[htp]
\caption{The corrections of Eqs.~(\ref{eq:B21D23D}-\ref{eq:B31D23D}), comparing between the coalescence factor obtained from a 1D and a 3D Gaussian fit to $\C_2$ computed in the BWM. We round the result to two significant digits.}
\begin{center}
\begin{tabular}{|c|c||c|c|c|c|c|c|}\hline
$p_t$&$m_t$&\;\;\;$\B_2^{1D}/\B_2^{3D}$\;\;\;&\;\;\;$\B_2^{1D}/\B_2^{3D}$\;\;\;&\;\;\;$\B_3^{1D}/\B_3^{3D}$\;\;\;&\;\;\;$\B_3^{1D}/\B_3^{3D}$\;\;\;&\;\;\;$\B_{3\Lambda}^{1D}/\B_{3\Lambda}^{3D}$\;\;\;&\;\;\;$\B_{3\Lambda}^{1D}/\B_{3\Lambda}^{3D}$\;\;\;\\
 ${\rm [GeV]}$ &${\rm  [GeV]}$ & {\bf PbPb}\;\;\;&\;\;\;{\bf pp}\;\;\;&\;\;\; {\bf PbPb}\;\;\;&\;\;\; {\bf pp}\;\;\;&\;\;\; {\bf PbPb}\;\;\;&\;\;\; {\bf pp}\;\;\;\\\hline\hline
0&0.94&1&1&1&1&1&1\\
0.5&1.07&1&1&1&1&1&1\\
1&1.37&1&1.1&1&1.1&1&1.1\\
2&2.2&1.1&1.3&1.2&1.9&1.2&1.5\\\hline
\end{tabular}\label{tab:B2corr}
\end{center}
\label{default}
\end{table}

We have also done a numerical calculation of the ratio $\B_2^{1D}/\B_2^{3D}$ using the Hulthen D wave function. In this exercise we calculated $\B_2^{3D}$ from Eq.~(\ref{eq:B2}), comparing that with the result of Eq.~(\ref{eq:B2hul}). The results we find for PbPb and pp and for all values of $p_t$ are numerically very close to those found in Tab.~\ref{tab:B2corr} using the Gaussian wave function. 

The conclusion from Tab.~\ref{tab:B2corr} is that the 1D Gaussian parameterisation tends to slightly over-estimate the coalescence factor, in comparison to the more accurate 3D parameterisation. The effect is more pronounced in small systems than in PbPb, and increases with increasing $p_t$. Using the 1D $R_{\rm inv}$ parameterisation results with $\mathcal{O}(10\%)$ error at $p_t\sim1$~GeV, rising to as much as a factor of 2 for $^3$He at $p_t=2$~GeV in pp collisions.  
It should be stressed that these conclusions are drawn from simple, model-dependent calculations in the BWM, and should only be taken as crude estimates of the theoretical uncertainty in the coalescence calculation.

\end{appendix}

\vspace{6 pt}

\bibliography{ref}

\end{document}